\documentstyle[thmsa,a4,sw20lart]{article}

\input{tcilatex}
\begin{document}

\author{S. Manoff \\
{\it Bulgarian Academy of Sciences,}\\
{\it Institute for Nuclear Research and Nuclear Energy,}\\
{\it Department of Theoretical Physics,}\\
{\it Blvd. Tzarigradsko Chaussee 72}\\
{\it 1784 Sofia - Bulgaria}}
\title{{\sc Mechanics of Continuous Media in }$(\overline{L}_n,g)${\sc -spaces. }\\
{\bf I. Introduction and mathematical tools}}
\date{e-mail address: smanov@inrne.bas.bg}
\maketitle

\begin{abstract}
Basic notions and mathematical tools in continuum media mechanics are
recalled. The notion of exponent of a covariant differential operator is
introduced and on its basis the geometrical interpretation of the curvature
and the torsion in $(\overline{L}_n,g)$-spaces is considered. The Hodge
(star) operator is generalized for $(\overline{L}_n,g)$-spaces. The
kinematic characteristics of a flow are outline in brief.

PACS numbers: 11.10.-z; 11.10.Ef; 7.10.+g; 47.75.+f; 47.90.+a; 83.10.Bb
\end{abstract}

\section{Introduction}

\subsection{Differential geometry and space-time geometry}

In the last years, the evolution of the relations between differential
geometry and space-time geometry has made some important steps toward
applications of more comprehensive differential-geometric structures in the
models of space-time than these used in (pseudo) Riemannian spaces without
torsion ($V_n$-spaces) \cite{Boothby}, \cite{Eddington}.

1. Recently, it has been proved that every differentiable manifold with one
affine connection and metrics [$(L_n,g)$-space] \cite{Hehl}, \cite
{Barvinskii} could be used as a model for a space-time. In it, the
equivalence principle (related to the vanishing of the components of an
affine connection at a point or on a curve in a manifold) holds \cite
{Iliev-1} $\div $ \cite{Iliev-7}, \cite{Hartley}. Even if the manifold has
two different (not only by sign) connections for tangent and co-tangent
vector fields [$(\overline{L}_n,g)$-space] \cite{Manoff-0}, \cite{Manoff-01}
the principle of equivalence is fulfilled at least for one of the two types
of vector fields \cite{Manoff-1}. On this grounds, every free moving
spinless test particle in a suitable basic system (frame of reference) \cite
{Manoff-2}, \cite{Manoff-3} will fulfil an equation identical with the
equation for a free moving spinless test particle in the Newtons mechanics
or in the special relativity. In $(\overline{L}_n,g)$- and $(L_n,g)$-spaces,
this equation could be the auto-parallel equation [different from the
geodesic equation in contrast to the case of (pseudo) Riemannian spaces
without torsion ($V_n$-spaces)].

2. There are evidences that $(L_n,g)$- and $(\overline{L}_n,g)$-spaces can
have similar structures as the $V_n$-spaces for describing dynamical systems
and the gravitational interaction. In such type of spaces one could use
Fermi-Walker transports \cite{Manoff-4} $\div $ \cite{Manoff-6} conformal
transports \cite{Manoff-7}, \cite{Manoff-8} and different frames of
reference \cite{Manoff-3}. All these notions appear as generalizations of
the corresponding notions in $V_n$-spaces. For instance, in $(L_n,g)$- and $(%
\overline{L}_n,g)$-spaces a proper non-rotating accelerated observer's frame
of reference could be introduced by analogy of the same type of frame of
reference related to a Fermi-Walker transport \cite{Misner} in the Einstein
theory of gravitation (ETG).

3. All kinematic characteristics related to the notion of relative velocity 
\cite{Stephani}, \cite{Kramer-2} as well as the kinematic characteristics
related to the notion of relative acceleration have been worked out for $%
(L_n,g)$- and $(\overline{L}_n,g)$-spaces without changing their physical
interpretation in $V_n$-spaces \cite{Manoff-9}. Necessary and sufficient
conditions as well as only necessary an only sufficient conditions for
vanishing shear, rotation and expansion acceleration induced by the
curvature are found \cite{Manoff-9}. The last results are related to the
possibility of building a theoretical basis for construction of
gravitational wave detectors on the grounds of gravitational theories over $%
(L_n,g)$ and $(\overline{L}_n,g)$--spaces.

Usually, in the gravitational experiments the measurements of two basic
objects are considered \cite{Grishchuk}:

(a) The relative velocity between two particles (or points) related to the
rate of change of the length (distance) between them. The change of the
distance is supposed to be caused by the gravitational interaction.

(b) The relative accelerations between two test particles (or points) of a
continuous media. These accelerations are related to the curvature of the
space-time and supposed to be induced by a gravitational force.

Together with the accelerations induced by the curvature in $V_n$-spaces
accelerations induced by the torsion would appear in $U_n$-spaces, as well
as by torsion and by non-metricity in $(L_n,g)$- and $(\overline{L}_n,g)$%
-spaces.

In particular, in other models of a space-time [different from the (pseudo)
Riemannian spaces without torsion] the torsion has to be taken into account
if we consider the characteristics of the space-time.

4. On the one hand, until now, there are a few facts that the torsion could
induce some very small and unmeasurable effects in quantum mechanical
systems considered in spaces with torsion \cite{Laemmerzahl}. At the same
time, there are no evidences that the model's descriptions of interactions
on macro level should include the torsion as a necessary mathematical tool.
From this (physical) point of view the influence of the torsion on dynamical
systems could be ignored since it could not play an important role in the
description of physical systems in the theoretical gravitational physics. On
the other hand, from mathematical point of view (as we will try to show in
this paper), the role of the torsion in new theories for description of
dynamical systems could be important and not ignorable.

\subsection{Differential geometry and continuous media mechanics}

1. In many considerations of the behavior of materials and physical systems
the motion of atoms and molecules in it is ignored and a material (physical
system) is considered as a whole subject. The molecular structure is not
investigated and the assumption is made that the matter is continuously
distributed in the volume occupied by it. This conception for a continuous
media appears as the basic conception of the mechanics of continuous media
(continuum media mechanics). In the range of the confined conditions, where
the basic conception could be valid, it could be used for description of
rigid bodies, fluids, and gases. Moreover, it has been shown \cite{Manoff-10}%
, \cite{Manoff-11} that every classical (non-quantized) field theory could
be considered as a theory of a continuous media. This fact gives rise to
generalization and application of the structures of continuous media
mechanics to different mathematical models of classical field theories and
especially in theories of gravitation.

2. In Einstein's theory of gravitation the relativistic continuous media
mechanics has been worked out by Synge \cite{Synge}, Lichnerowicz \cite
{Lichnerowicz} and especially by Ehlers \cite{Ehlers}. Most of the notions
of the classical mechanics of continuous media have been generalized and
applied to finding out invariant characteristics of Einstein's field
equations as well as to physical interpretation of formal results in the
gravitational theory. Hydrodynamics has also been considered as an essential
part of a gravitational theory \cite{Ehlers}.

3. Notions of relativistic continuous media mechanics have been generalized
for spaces with affine connections and metrics \cite{Manoff-9} $\div $ \cite
{Manoff-11}. Conditions have been found for these types of spaces admitting
special vector fields with vanishing kinematic characteristics related to
the notions of mechanics of continuous media. Nevertheless, the physical
interpretation of the generalized notions is in some cases not so obvious as
in the classical continuous mechanics and requires additional
considerations. Usually, this is done by the use of congruencies (families
of non-intersecting curves), interpreted as world lines describing the
motion of matter elements in an infinitesimal region of a flow \cite{Ehlers}%
, \cite{Stephani}. The general considerations include determination of the
kinematic characteristics of a flow such as the velocity, the relative
velocity and relative acceleration between the material points (elements) as
well as the influence of the geometric structure (curvature, torsion) on
these kinematic characteristics. An invariant description of a continuous
media is related to the notions of covariant derivatives along curves, to
covariant and Lie differential operators acting on corresponding vector and
tensor fields. The acceptance of the hypothesis of a continuous media
(continuum) as a basis for mathematical description of the behavior of
matter means also that the set of quantities (such as deformation, stress,
transports, draggings along etc.) should be expressed by means of (at least)
piecewise continuous functions of space and time.

In the present paper, basic notions of continuum media mechanics are
introduced and considered in spaces with affine connections and metrics [$(%
\overline{L}_n,g)$-spaces]. In Section 2 basic notions and mathematical
tools are recalled. The notion of exponent of a covariant differential
operator is introduced and on its basis the geometrical interpretation of
the curvature and the torsion in $(\overline{L}_n,g)$-spaces is considered
in Section 3 and Section 4. In Section 5 the kinematic characteristics of a
flow are considered.

All considerations are given in details (even in full details) for those
readers who are not familiar with the considered problems.

{\it Remark.} The present paper is the first part of a larger research
report on the subject with the title ''Contribution to continuous media
mechanics in $(\overline{L}_n,g)$-spaces'' and with the following contents:

I. Introduction and mathematical tools.

II. Relative velocity and deformations.

III. Relative accelerations.

IV. Stress (tension) tensor.

The parts are logically self-dependent considerations of the main topics
considered in the report.

\section{Basic notions and mathematical tools}

For the further considerations we need to recall some well known facts and
definitions from continuous media mechanics.

A mathematical model of a continuum (continuous media) is assumed to be a
differentiable manifold $M$ with dimension $\dim M=n$. In this continuum all
structures, kinematic and dynamic characteristics of a media could be
described and considered by the use of all differential-geometric structures
in the differentiable manifold $M$. The basic notions of continuous media
mechanics are related to problems of description of \cite{Mase}

(a) deformations

(b) stresses (tensions)

(c) relation between stresses (tensions) and deformations

(d) dynamical reasons for generation of deformations and stresses (tensions).

The notion of motion and flow are used in the sense of instant (in a moment)
motion and in the sense of continuous motion. In some cases the notion of
flow is used in the sense of motion causing residual deformation. In the
theory of fluids (as a part of continuous media mechanics) the notion of
flow means continuous motion. We will use the notion of flow (when no other
conditions are supposed) in the sense of continuous motion.

The reason for deformation in a media is the motion of its material points
to each other. This motion could be expressed in two different ways:

(a) By means of the change of a vector field $\xi $ (called deformation or
deviation vector field) along a vector field $u$, interpreted as the
velocity of the material points, i.e. a deformation is described by means of
the covariant derivative $\nabla _u\xi $ of $\xi $ along $u$.

(b) By means of the change of the velocity vector field $u$ along the
deformation (deviation) vector field $\xi $, i.e. a deformation is described
by means of the covariant derivative $\nabla _\xi u$ of $u$ along $\xi $.

The vector fields $u$ and $\xi $ could be chosen as tangent vector fields to
the co-ordinates introduced in a media, considered as a differentiable
manifold $M$ with $\dim M=n$. For $n\leq 3$ the classical (non-relativistic)
continuous media mechanics has been developed. For $n=4$ the relativistic
continuous media mechanics has been worked out. For $n\geq 4$ models could
also be worked out describing a continuous media in more sophisticated
spaces with affine connections and metrics.

1. There are basic definitions related to the notion of flow in the
continuous media mechanics. Let us recall some of them.

{\it Trajectory} is a line at which a material point moves during its motion.

{\it Motion} is the change of the position of a material point in a
differentiable manifold considered as a model of space-time.

{\it Position} of a material point is the point of a differentiable manifold
identified with the material point.

{\it Space} is a differentiable manifold provided with additional structures
and considered as a model of a physical space or space-time.

{\it Physical space} is a sub space of a space-time considered in many
mathematical models of space-time as a sub space orthogonal to the time.

{\it Homogeneous} are spaces or matter if they have the same properties at
all points of a manifold (considered as a model of a space, space-time, or
matter).

{\it Isotropic} are spaces or matter with respect to a given property if
this property at a point of a manifold is the same for all directions going
out of this point.

{\it Anisotropic} are spaces or matter with respect to a given property if
this property is different for different directions going out of a given
point.

{\it Line of a flow} is a line with tangential vector at each of its points
collinear (or identical) with the velocity of a material point with the
corresponding position.

{\it Flow} is a congruence (family) of flow's lines.

2. Every motion of material points of a dynamical system is described by the
use of a frame of reference \cite{Manoff-2} related to an observer with
velocity vector field $u$.

The use of a contravariant non-isotropic (non-null) vector field $u$ and its
corresponding projective metrics $h^u$\thinspace and $h_u$ is analogous to
the application of a non-isotropic (time-like) vector field in the s. c. 
{\it monad formalism} [$(3+1)\,$-{\it decomposition}] in $V_4$-spaces for
description of dynamical systems in Einstein's theory of gravitation (ETG) 
\cite{Eckart} $\div $\cite{Schmutzer-2}.

The contravariant time-like vector field $u$ has been interpreted as a
tangential vector field at the world line of an observer determining the
frame of reference (the reference frame) in the space-time. By the use of
this reference frame a given physical system is observed and described. The
characteristics of the vector field determine the properties of the
reference frame. Moreover, the vector field is assumed to be an absolute
element in the scheme for describing the physical processes, i. e. the
vector field is not an element of the model of the physical system. It is
introduced as {\it a priory} given vector field which does not depend on the
Lagrangian system. In fact, the physical interpretation of the contravariant
non-isotropic vector field $u$ can be related to two different approaches
analogous to the method of Lagrange and the method of Euler in describing
the motion of liquids in the hydrodynamics \cite{Pavlenko}.

The $(n-1)+1$ decomposition (monad formalism) can be related to the method
of Euler or to the method of Lagrange on the basis of a frame of reference.

There are at least two types of frame of reference with respect to the
motion of material points:

(a) {\it Proper frame of reference}. A frame of reference of an observer,
moving with a material point of the dynamical system, is called proper frame
of reference for this point. The position and the velocity of the observer
is identified with the position and the velocity of the material point. Its
velocity is determined by means of the dynamics of the flow, described by
the use of equations of the type of Euler-Lagrange's equations. Solutions of
these equations with some initial data describe the motion of the material
point (observer) with respect to a given initial time. So, a proper frame of
reference is related to the method of Lagrange in continuous media mechanics.

In the method of Lagrange, the object with the considered motion appears as
a point (particle) of a flow. The motion of this point is given by means of
equations for a vector field $u$ interpreted as the velocity of the
particle. The solutions of these equations give the trajectories of the
points in the flow as basic characteristics of the physical system. Here,
the vector field $u$ appears as an element of the model of the system. It is
connected with the motions of the system's elements. Therefore, a Lagrangian
system (and respectively its Lagrangian density) could contain as an
internal characteristic a contravariant vector field $u$ obeying equations
of the type of the Euler-Lagrange equations and describing the evolution of
the system.

{\it Lagrangian system} is a dynamical system (system obeying physical lows)
which characteristics are described by means of an Lagrangian density and
its corresponding Euler-Lagrange's equations.

(a$_1$) {\it Method of Lagrange}. The vector field $u$ is interpreted as the
velocity vector field of a continuum media with an observer co-moving with
it . The last assumption means that the velocity vector field of the
observer is identical with the velocity vector field of the media where he
is situated.

The motion of the observer (his velocity vector field) is determined by the
characteristics of the (Lagrangian) system. Its velocity vector field is, on
the other side, determined by the dynamical characteristics of the system by
means of equations of the type of the Euler-Lagrange equations.

(b) {\it Non-proper frame of reference}. A frame of reference of an
observer, moving in space-time with his own velocity and describing the
motion of material points by means of projections of the characteristics of
a flow on its own kinematic characteristics of the motion, is called
non-proper frame of reference.

The position and the velocity of the observer could be different from the
position and the velocity of the observed material point. The equations
describing the motions of the observer and the material points could be
different. The equation of motion of the observer is assumed to be given.

The non-proper frame of reference is related to the method of Euler in
continuous media mechanics.

(b$_1$) {\it Method of Euler}. In the method of Euler, the object with the
considered motion appears as the model of the continuum media. Instead of
the investigation of the motion of every (fixed by its velocity and
position) point (particle), the kinematic characteristics in every immovable
point in the space are considered as well as the change of these
characteristics after moving on to an other space point. The motion is
assumed to be described if the vector field $u$ is considered as a given (or
known) velocity vector field.

The vector field $u$ is interpreted as the velocity vector field of an
observer who describes a physical system with respect to his vector field
(his velocity). This physical system is characterized by means of a
Lagrangian system (Lagrangian density).

The motion of the observer (his velocity vector field) is given
independently of the motion of the considered Lagrangian system.

In a proper frame of reference all kinematic characteristics of the vector
field $u$ are at the same time kinematic characteristics of the flow. In a
non-proper frame of reference the kinematic characteristics of the vector
field $u$ are only characteristics of the frame of reference and not
characteristics of the flow.

In our further considerations we will assume the existence of a proper frame
of reference in a flow.

3. Continuous media mechanics describes the state of a continuum by the use
or two major characteristics (deformation and stress) and relations between
them.

(a) A deformation in a continuous media describes the change of the relative
positions of the material points in the media. Its invariant (tensor)
characteristic is the deformation velocity tensor and the related to it
kinematic characteristics called shear velocity tensor, rotation (vortex)
velocity tensor, and expansion velocity invariant \cite{Manoff-9}.

(b) A stress (tension) in a continuous media describes the surface forces in
the media. A {\it surface force} is a force acting on a surface element (an
element of a surface covering an invariant volume element of continuous
media). It is then related to a volume force. A {\it volume force }is a
force acting on a {\it invariant volume element of the media} identified
with the invariant volume element in a differentiable manifold $M$.

\subsection{Covariant derivative of a tensor field along a curve. Exponent
of the covariant differential operator}

The notion of covariant derivative was introduced as the result of the
action of the covariant differential operator on a given tensor field (with
finite rank). The covariant differential operator is defined as a
differential operator (along a given contravariant vector field) determining
the corresponding affine connection. On the other side, a given
contravariant vector field $u$ could be considered as a tangent vector field
along a parametrized curve $x^i(\tau )$, $i=1,2,...,n$, $\tau \in {\bf R}$,
in the manifold $M$ and written in the form 
\begin{equation}
u:=\frac d{dt}=\frac{dx^i}{d\tau }\cdot \partial _i=u^i\cdot \partial _i%
\text{ , \ }u^i=\frac{dx^i(\tau )}{d\tau }\text{ \ \ .}  \label{1.1}
\end{equation}

The covariant derivative of a tensor field $V=V^A\,_B\cdot \partial
_A\otimes dx^B$, $V\in \otimes ^k\,_l(M)$ along the vector field $u$%
\begin{equation}
\nabla _uV=V^A\,_{B;i}\cdot u^i\cdot \partial _A\otimes dx^B  \label{1.2}
\end{equation}

\noindent %
for $u=d/d\tau $ is usually written in the form 
\begin{equation}
\nabla _uV=\nabla _{\frac d{d\tau }}V:=\frac{DV}{d\tau }\,\ \ \ \text{.}
\label{1.3}
\end{equation}

At the point of the curve $x^i(\tau _0=\,$const.$,\lambda _0^a=\,$const.$)$
with the parameter $\tau _0=$const. the covariant derivative could be
written as 
\begin{equation}
\left( \nabla _uV\right) _{(\tau _0)}=(\nabla _{\frac d{d\tau }}V)_{(\tau
_0)}=\left( \frac{DV}{d\tau }\right) _{(\tau _0)}\text{ , \ \ \ \ \ }\lambda
_0^a=\text{const.},  \label{1.4}
\end{equation}

\noindent %
where $V_{(x^{i}(\tau ,\lambda _{0}^{a}))}=V_{(\tau )}$.

The question arises how can we express the covariant derivative of \ $%
V(x^i(\tau ))=V(\tau )$ at the point $x^i(\tau _0+dt)$, $d\tau =\varepsilon
\ll 1$, \ by means of the covariant derivative at the point $x^i(\tau _0)$.

\paragraph{Exponential mapping of an ordinary differential operator}

If we wish to express only the components $V^A\,_B(x^i(\tau ))=V^A\,_B(\tau
)\in C^r(M)$ at the point $x^i(\tau _0+\varepsilon )$ by means of the
components $V^A\,_B(x^i(\tau _0))=V^A\,_B(\tau _0)$ we can use the
decomposition of $V^A\,_B(\tau _0+\varepsilon )$ in Taylor raw with respect
to $V^A\,_B(\tau _0)$ in the form 
\begin{eqnarray}
V^A\,_B(\tau _0+\varepsilon ) &=&V^A\,_B(\tau _0)+\varepsilon \cdot \left( 
\frac{dV^A\,_B}{d\tau }\right) _{(\tau _0)}+\frac 1{2!}\cdot \varepsilon
^2\cdot \left( \frac{d^2V^A\,_B}{d\tau ^2}\right) _{(\tau _0)}+\cdots = 
\nonumber \\
&=&\left[ (1+\varepsilon \cdot \frac d{d\tau }+\frac 1{2!}\cdot \varepsilon
^2\cdot \frac{d^2}{d\tau ^2}+\cdots )V^A\,_B\right] _{(\tau _0)}=  \nonumber
\\
&=&\left[ \left( \exp [\varepsilon \cdot \frac d{d\tau }]\right)
V^A\,_B\right] _{(\tau _0)}\text{ .}  \label{1.5}
\end{eqnarray}

The operator 
\begin{equation}
\exp [\varepsilon \cdot \frac d{d\tau }]=1+\varepsilon \cdot \frac d{d\tau
}+\frac 1{2!}\cdot \varepsilon ^2\cdot \frac{d^2}{d\tau ^2}+\cdots
\label{1.6}
\end{equation}

\noindent %
is called {\it exponent of the ordinary differential operator} $\varepsilon
\cdot \frac d{d\tau }$ \cite{Schutz}. It maps the components $V^A\,_B(\tau
_0)$ at a given point $x^i(\tau _0)$ of the curve with a parameter $\tau _0$
to components $V^A\,_B(\tau _0+d\tau )$ at the point $x^i(\tau )$ of the
curve with parameter $\tau =$ $\tau _0+d\tau $, where $d\tau =\varepsilon
\ll 1$. The exponent mapping of $V^A\,_B(\tau _0)$ into $V^A\,_B(\tau
_0+d\tau )$ has as a precondition the conservation (not changing) of the
tensor bases $\partial _A\otimes dx^B$ on the curve $x^i(\tau )$. This means
that we are considering a tensor field $V=V^A\,_B\cdot \partial _A\otimes
dx^B$ in two different neighboring points at the curve assuming that the
tensor bases are one and the same at both the points. By the use of the
exponent of the ordinary differential operator we can find a geometrical
interpretation of the Lie derivative (the commutator of two contravariant
vector fields) of a contravariant vector field along an other contravariant
vector field \cite{Schutz}.

\subsubsection{Geometrical interpretation of the Lie derivative of a
contravariant vector field along an other contravariant vector field}

Let a two parametric congruence of curves $x^i(\tau ,\lambda )$ be given. An
infinitesimal quadrangle is determined by its apexes $ABDC$ (the points $A$, 
$B$, $D$, $C$) with the co-ordinates respectively: 
\begin{eqnarray}
A &:&(\tau _0,\lambda _0)\text{ \thinspace \thinspace \thinspace \thinspace
, \thinspace \thinspace \thinspace \thinspace \thinspace \thinspace
\thinspace \thinspace \thinspace \thinspace \thinspace \thinspace \thinspace
\thinspace \thinspace \thinspace \thinspace \thinspace \thinspace \thinspace
\thinspace \thinspace \thinspace \thinspace \thinspace \thinspace \thinspace
\thinspace \thinspace \thinspace \thinspace \thinspace \thinspace \thinspace
\thinspace \thinspace \thinspace \thinspace \thinspace \thinspace \thinspace
\thinspace \thinspace \thinspace \thinspace \thinspace \thinspace \thinspace
\thinspace \thinspace \thinspace \thinspace \thinspace \thinspace \thinspace
\thinspace \thinspace \thinspace \thinspace \thinspace \thinspace \thinspace
\thinspace \thinspace \thinspace \thinspace \thinspace \thinspace \thinspace
\thinspace \thinspace }B:(\tau _0,\lambda _0+d\lambda )\text{ \thinspace
\thinspace \thinspace \thinspace \thinspace ,\thinspace \thinspace
\thinspace \thinspace }  \nonumber \\
\text{\thinspace \thinspace \thinspace \thinspace \thinspace }D &:&(\tau
_0+k_1\cdot d\tau ,\lambda _0+k_2\cdot d\lambda )\text{ \thinspace
\thinspace \thinspace \thinspace \thinspace \thinspace \thinspace
,\thinspace \thinspace \thinspace \thinspace \thinspace \thinspace
\thinspace \thinspace \thinspace \thinspace \thinspace \thinspace \thinspace
\thinspace \thinspace \thinspace \thinspace \thinspace \thinspace \thinspace
\thinspace \thinspace \thinspace \thinspace }C:(\tau _0+d\tau ,\lambda _0)%
\text{ \thinspace \thinspace \thinspace \thinspace \thinspace \thinspace ,} 
\nonumber \\
k_1,k_2 &\in &{\bf R}\text{ .}  \label{1.7}
\end{eqnarray}

Let us assume that $d\tau =d\lambda =\varepsilon \ll 1$, and $k_1\cdot d\tau
=k_2\cdot d\lambda =k\cdot \varepsilon \ll 1$. We can now represent the
co-ordinates of the points lying at equal parameters $\varepsilon $ from the
points $A$, $B$, and $C$ by the use of the co-ordinates of the point $A$ and
the exponent of the ordinary differential operator. Let these points be $C_1$
and $D_1$ with co-ordinates $C_1:$ $(\tau _0+d\tau ,\lambda _0+d\lambda )$
on the way $ACC_1$ and $D_1:(\tau _0+d\tau ,\lambda _0+d\lambda )$ on the
way $ABD_1$. The points $A$, $B$, $C$, $D$, $C_1$, and $D_1$ will have the
co-ordinates expressed by the co-ordinates of the point $A$ as follows: 
\begin{eqnarray}
A &:&x^i(\tau _0,\lambda _0)  \nonumber \\
B &:&x^i(\tau _0,\lambda _0+d\lambda )=\left[ \left( \exp [d\lambda \cdot
\frac d{d\lambda }]\right) x^i\right] _{(\tau _0,\lambda _0)}\text{ ,} 
\nonumber \\
C &:&x^i(\tau _0+d\tau ,\lambda _0)=\left[ \left( \exp [d\tau \cdot \frac
d{d\tau }]\right) x^i\right] _{(\tau _0,\lambda _0)}\text{ \thinspace ,} 
\nonumber \\
D &:&x^i(\tau _0+k_1\cdot d\tau ,\lambda _0+k_2\cdot d\lambda )=\left( \exp
[k_1\cdot d\tau \cdot \frac d{d\tau }]\right) x^i\,_{(\tau _0,\lambda
_0+d\lambda )}=  \nonumber \\
&=&\left[ \left( \exp [k_1\cdot d\tau \cdot \frac d{d\tau }]\right) \circ
\left( \exp [k_2\cdot d\lambda \cdot \frac d{d\lambda }]\right) x^i\right]
_{(\tau _0,\lambda _0)}  \nonumber \\
C_1 &:&x^i(\tau _0+d\tau ,\lambda _0+d\lambda )=\left( \exp [d\lambda \cdot
\frac d{d\lambda }]\right) x_{_{(\tau _0+d\tau ,\lambda _0)}}^i=  \nonumber
\\
&=&\,\left[ \left( \exp [d\lambda \cdot \frac d{d\lambda }]\right) \circ
\left( \exp [d\tau \cdot \frac d{d\tau }]\right) x^i\right] _{(\tau
_0,\lambda _0)}\text{ \thinspace ,}  \nonumber \\
D_1 &=&x^i(\tau _0+d\tau ,\lambda _0+d\lambda )=\left( \exp [d\tau \cdot
\frac d{d\tau }]\right) x^i\,_{(\tau _0,\lambda _0+d\lambda )}=  \nonumber \\
&=&\left[ \left( \exp [d\tau \cdot \frac d{d\tau }]\right) \circ \left( \exp
[d\lambda \cdot \frac d{d\lambda }]\right) x^i\right] _{(\tau _0,\lambda _0)}%
\text{ \thinspace \thinspace \thinspace \thinspace .}  \label{1.8}
\end{eqnarray}

Now, we have to consider the co-ordinates of the points $C_1$ and $D_1$. For 
$d\tau =d\lambda =\varepsilon $ we have 
\begin{eqnarray}
C_1 &:&x^i(\tau _0+\varepsilon ,\lambda _0+\varepsilon )=  \nonumber \\
&=&\,\left[ \left( \exp [\varepsilon \cdot \frac d{d\lambda }]\right) \circ
\left( \exp [\varepsilon \cdot \frac d{d\tau }]\right) x^i\right] _{(\tau
_0,\lambda _0)}\text{ \thinspace ,}  \nonumber \\
D_1 &:&x^i(\tau _0+\varepsilon ,\lambda _0+\varepsilon )=  \nonumber \\
&=&\left[ \left( \exp [\varepsilon \cdot \frac d{d\tau }]\right) \circ
\left( \exp [\varepsilon \cdot \frac d{d\lambda }]\right) x^i\right] _{(\tau
_0,\lambda _0)}\text{ \thinspace \thinspace \thinspace \thinspace .}
\label{1.9}
\end{eqnarray}

By the use of the explicit form of the exponent of the ordinary differential
operators $d/d\tau $ and $d/d\lambda $ (identical with the tangent vector
fields $u$ and $\xi $ at the congruence of curves), we obtain 
\begin{eqnarray}
C_1 &:&\,\stackunder{AC}{x^i}(\tau _0+\varepsilon ,\lambda _0+\varepsilon )=
\nonumber \\
&=&\left[ (1+\varepsilon \cdot \frac d{d\lambda }+\frac 1{2!}\cdot
\varepsilon ^2\cdot \frac{d^2}{d\lambda ^2}+\cdots )\circ (1+\varepsilon
\cdot \frac d{d\tau }+\frac 1{2!}\cdot \varepsilon ^2\cdot \frac{d^2}{d\tau
^2}+\cdots )x^i\right] _{(\tau _0,\lambda _0)}\text{ ,}  \nonumber \\
D_1 &:&\stackunder{AB}{x^i}(\tau _0+\varepsilon ,\lambda _0+\varepsilon )= 
\nonumber \\
&=&\left[ (1+\varepsilon \cdot \frac d{d\tau }+\frac 1{2!}\cdot \varepsilon
^2\cdot \frac{d^2}{d\tau ^2}+\cdots )\circ (1+\varepsilon \cdot \frac
d{d\lambda }+\frac 1{2!}\cdot \varepsilon ^2\cdot \frac{d^2}{d\lambda ^2}%
+\cdots )x^i\right] _{(\tau _0,\lambda _0)}\text{ \thinspace .}  \label{1.10}
\end{eqnarray}

The difference between the co-ordinates $\stackunder{AC}{x^i}(\tau
_0+\varepsilon ,\lambda _0+\varepsilon )$ and $\stackunder{AB}{x^i}(\tau
_0+\varepsilon ,\lambda _0+\varepsilon )$ can be found as 
\begin{eqnarray}
&&\stackunder{AC}{x^i}(\tau _0+\varepsilon ,\lambda _0+\varepsilon )-\,%
\stackunder{AB}{x^i}(\tau _0+\varepsilon ,\lambda _0+\varepsilon )  \nonumber
\\
&=&\left[ \left( \exp [\varepsilon \cdot \frac d{d\lambda }]\right) \circ
\left( \exp [\varepsilon \cdot \frac d{d\tau }]\right) x^i\right] _{(\tau
_0,\lambda _0)}-\left[ \left( \exp [\varepsilon \cdot \frac d{d\tau
}]\right) \circ \left( \exp [\varepsilon \cdot \frac d{d\lambda }]\right)
x^i\right] _{(\tau _0,\lambda _0)}  \nonumber \\
&=&\left[ \left( \exp [\varepsilon \cdot \frac d{d\lambda }]\right) ,\left(
\exp [\varepsilon \cdot \frac d{d\tau }]\right) \right] x_{(\tau _0,\lambda
_0)}^i\text{ .}  \label{1.11}
\end{eqnarray}

Up to the second order of $\varepsilon $, we have 
\begin{eqnarray}
C_1 &:&\,\stackunder{AC}{x^i}(\tau _0+\varepsilon ,\lambda _0+\varepsilon )=
\nonumber \\
&=&\left[ [1+\varepsilon \cdot (\frac d{d\lambda }+\frac d{d\tau
})+\varepsilon ^2\cdot (\frac 12\cdot \frac{d^2}{d\lambda ^2}+\frac
d{d\lambda }\circ \frac d{d\tau }+\frac 12\cdot \frac{d^2}{d\tau ^2}%
)+O(\varepsilon ^3)]x^i\right] _{(\tau _0,\lambda _0)}\text{\thinspace
\thinspace ,}  \nonumber \\
D_1 &:&\stackunder{AB}{x^i}(\tau _0+\varepsilon ,\lambda _0+\varepsilon )= 
\nonumber \\
&=&\left[ [1+\varepsilon \cdot (\frac d{d\tau }+\frac d{d\lambda
})+\varepsilon ^2\cdot (\frac 12\cdot \frac{d^2}{d\tau ^2}+\frac d{d\tau
}\circ \frac d{d\lambda }+\frac 12\cdot \frac{d^2}{d\lambda ^2}%
)+O(\varepsilon ^3)]x^i\right] _{(\tau _0,\lambda _0)}\text{ ,}  \nonumber \\
&&\stackunder{AC}{x^i}(\tau _0+\varepsilon ,\lambda _0+\varepsilon )-\,%
\stackunder{AB}{x^i}(\tau _0+\varepsilon ,\lambda _0+\varepsilon )  \nonumber
\\
&=&\left\{ \left[ \varepsilon ^2\cdot [\frac d{d\lambda }\circ \frac d{d\tau
}-\frac d{d\tau }\circ \frac d{d\lambda }]+O(\varepsilon ^3)\right]
x^i\right\} _{(\tau _0,\lambda _0)}\text{\thinspace }  \nonumber \\
&=&\left[ \varepsilon ^2\cdot [\xi \circ u-u\circ \xi ]+O(\varepsilon
^3)\right] _{(\tau _0,\lambda _0)}\approx \varepsilon ^2\cdot \left( [\xi
,u]x^i\right) _{(\tau _0,\lambda _0)}\text{ \thinspace \thinspace \thinspace
\thinspace .}  \label{1.12}
\end{eqnarray}

On this basis, the relation 
\begin{eqnarray}
\stackunder{AC}{x^i}-\stackunder{AB}{x^i} &\approx &\varepsilon ^2\cdot
\left( [\frac d{d\lambda }\circ \frac d{d\tau }-\frac d{d\tau }\circ \frac
d{d\lambda }]x^i\right) _{(\tau _0,\lambda _0)}=  \nonumber \\
&=&\varepsilon ^2\cdot \left( [\xi ,u]x^i\right) _{(\tau _0,\lambda
_{0)}}=\varepsilon ^2\cdot \left[ (\pounds _\xi u)x^i\right] _{(\tau
_0,\lambda _0)}\text{ \thinspace \thinspace \thinspace \thinspace }
\label{1.13}
\end{eqnarray}

\noindent follows.

Therefore, up to the second order of $\varepsilon $, the Lie derivative $%
\pounds _\xi u=[\xi ,u]=-\pounds _u\xi $ could be interpreted as a measure
for the difference between the co-ordinates of two points $C_1$ and $D_1$
lying at equal distances (parameters) from a starting point $A$. This means
that if a (material) point is moving from a point $A$ across a point $B$ to
a point $D_1$ (with equal distances $\varepsilon $ from point $A$ to point $%
B $ and from point $B$ to point $D_1$), it could not meet a point, moving
from a point $A$ across a point $C$ to a point $C_1$ (with equal distances $%
\varepsilon $ between point $A$ and point $C$ and from point $C$ to point $%
C_1$), if the Lie derivative $\pounds _\xi u=-\pounds _u\xi $ of the tangent
vector $\xi $ along $u$ or vice versa is not equal to zero. If $\pounds _\xi
u=-\pounds _u\xi =0$ then both the points $C_1$ and $D_1$ will coincide and
the points, moving on the different paths $ACC_1$ and $ABD_1$ will meet at
one and the same point $D_1\equiv C_1$ closing the quadrangle $AB(C_1\equiv
D_1)C$. In the opposite case, $ABC_1D_1C$ will construct a pentagon.

On the basis of the relation 
\begin{equation}
\pounds _\xi u=\nabla _\xi u-\nabla _u\xi -T(\xi ,u)\text{ , }  \label{1.14}
\end{equation}

\noindent %
under the conditions for parallel transport of $\xi $ along $u$ and $u$
along $\xi $, i.e. under the conditions $\nabla _u\xi =0$ and $\nabla _\xi
u=0$, it follows that 
\begin{equation}
\pounds _\xi u=-T(\xi ,u)\text{ \thinspace \thinspace \thinspace .}
\label{1.15}
\end{equation}

The last relation is used for finding another geometric interpretation of
the torsion vector $T(\xi ,u)$ and respectively of the torsion tensor $T$.
Under the conditions $\nabla _u\xi =0$ and $\nabla _\xi u=0$ the torsion
vector (and the Lie derivative) appear as a measure for the non-existence of
a closed infinitesimal quadrangle with equal sides. This fact is used in
theories of crystals for description of the defects in crystal's cells.

\subsubsection{Exponent of the covariant differential operator}

The covariant operator $\frac D{d\tau }$ is a generalization of the
differential operator $\frac d{d\tau }$ along the curve $x^i(\tau )$ for the
case when not only the components $V^A\,_B(\tau )$ of the tensor field $%
V(\tau )=V^A\,_B(\tau )\cdot \partial _A(\tau )\otimes dx^B(\tau )$ are
changing along the curve but also its tensor bases $\partial _A(\tau
)\otimes dx^B(\tau )$ are also changing along the curve in accordance with
the relations 
\begin{eqnarray}
\frac D{d\tau }(\partial _A\otimes dx^B) &=&\nabla _u(\partial _A\otimes
dx^B)=(\Gamma _{Ai}^C\cdot \partial _C\otimes dx^B+\partial _A\otimes
P_{Di}^B\cdot dx^D)\cdot u^i=  \nonumber \\
&=&u^i\cdot (\Gamma _{Ai}^C\cdot g_D^B\cdot \partial _C\otimes
dx^D+P_{Di}^B\cdot g_A^C\cdot \partial _C\otimes dx^D)=  \nonumber \\
&=&u^i\cdot (\Gamma _{Ai}^C\cdot g_D^B+P_{Di}^B\cdot g_A^C)\cdot \partial
_C\otimes dx^D=  \nonumber \\
&=&\frac{dx^i}{d\tau }\cdot (\Gamma _{Ai}^C\cdot g_D^B+P_{Di}^B\cdot
g_A^C)\cdot \partial _C\otimes dx^D\text{ \ ,}  \label{1.16}
\end{eqnarray}

\noindent where 
\begin{eqnarray*}
\Gamma _{Aj}^B &=&-S_{Am}\,^{Bi}.\Gamma _{ij}^m\text{ ,\thinspace \thinspace
\thinspace \thinspace \thinspace \thinspace \thinspace \thinspace \thinspace
\thinspace \thinspace \thinspace \thinspace }A=j_1...j_l\text{ ,\thinspace
\thinspace \thinspace }B=i_1...i_l\text{ ,\thinspace } \\
S_{Am}\,^{Bi}
&=&-%
\sum_{k=1}^lg_{j_k}^i.g_m^{i_k}.g_{j_1}^{i_1}.g_{j_2}^{i_2}...g_{j_{k-1}}^{i_{k-1}}.g_{j_{k+1}}^{i_{k+1}}...g_{j_l}^{i_l}%
\text{ \thinspace \thinspace ,\thinspace } \\
P_{Aj}^B &=&-S_{Am}^{Bi}.P_{ij}^m\,\,\,\text{,} \\
g_B^A
&=&g_{i_1}^{j_1}...g_{i_{m-1}}^{j_{m-1}}.g_{i_m}^{j_m}.g_{i_{m+1}}^{j_{m+1}}...g_{i_l}^{j_l}%
\text{ \thinspace \thinspace \thinspace \thinspace .}
\end{eqnarray*}

Only if the components $\Gamma _{jk}^i$ and $P_{jk}^i$ of the contravariant
and covariant affine connections $\Gamma $ and $P$ respectively vanish along
the curve, then 
\begin{equation}
\frac D{d\tau }(\partial _C\otimes dx^D)=0  \label{1.17}
\end{equation}

\noindent %
and the tensor bases do not change. Therefore, if we consider the tensor $%
V(\tau _0+d\tau )$ at the point $x^i(\tau _0+d\tau )=x^i(\tau _0+\varepsilon
)$ of the curve and wish to compare it with the tensor $V(\tau _0)$ at the
point $x^i(\tau _0)$ of the same curve in a space with a contravariant
affine connection $\Gamma $ and a covariant affine connection $P$ we should
change the operator $\frac d{d\tau }$ with the operator $\frac D{d\tau }$ in
the representation of $V(\tau _0+\varepsilon )=V_{(\tau _0+\varepsilon )}$
by the use of $V(\tau _0)=V_{(\tau _0)}$ and its covariant derivatives along
the curve $x^i(\tau )$, i.e. 
\begin{eqnarray}
V_{(\tau _0+\varepsilon )} &=&V_{(\tau _0)}+\varepsilon \cdot \left( \frac{DV%
}{d\tau }\right) _{(\tau _0)}+\frac 1{2!}\cdot \varepsilon ^2\cdot \left( 
\frac{D^2V}{d\tau ^2}\right) _{(\tau _0)}+\cdots =  \nonumber \\
&=&\left[ (1+\varepsilon \cdot \frac D{d\tau }+\frac 1{2!}\cdot \varepsilon
^2\cdot \frac{D^2}{d\tau ^2}+\cdots )V\right] _{(\tau _0)}=  \nonumber \\
&=&\left[ \left( \exp [\varepsilon \cdot \frac D{d\tau }]\right) V\right]
_{(\tau _0)}\text{ .}  \label{1.18}
\end{eqnarray}

The operator 
\begin{equation}
\exp [\varepsilon \cdot \frac D{d\tau }]=1+\varepsilon \cdot \frac D{d\tau
}+\frac 1{2!}\cdot \varepsilon ^2\cdot \frac{D^2}{d\tau ^2}+\cdots
\label{1.19}
\end{equation}

\noindent %
could be called {\it exponent of the covariant differential operator} $%
\varepsilon \cdot \frac D{d\tau }$. It maps the tensor $V_{(\tau _0)}$ at
the point $x^i(\tau _0)$ of the curve $x^i(\tau )$ to the tensor $V_{(\tau
_0+\varepsilon )}$ at the point $x^i(\tau _0+\varepsilon )$ of the same
curve as 
\begin{equation}
V_{(\tau _0+\varepsilon )}=\left[ \left( \exp [\varepsilon \cdot \frac
D{d\tau }]\right) V\right] _{(\tau _0)}\text{ \ }  \label{1.20}
\end{equation}

\noindent %
under taking into account the change of the components of the tensor in a
(co-ordinate) basis as well as the change of the tensor bases along the
curve at two neighboring points of the curve. By the use of the exponent of
the covariant differential operator we can compare two tensors at two
neighboring points of a curve at one of its points.

The expression 
\begin{eqnarray}
\left[ \left( \exp [\varepsilon \cdot \frac D{d\tau }]\right) V\right]
_{(\tau =\tau _0+\varepsilon )} &=&\text{ }V_{(\tau =\tau _0)}+\varepsilon
\cdot \left( \frac{DV}{d\tau }\right) _{(\tau =\tau _0)}+  \nonumber \\
&&+\frac 1{2!}\cdot \varepsilon ^2\cdot \left( \frac{D^2V}{d\tau ^2}\right)
_{(\tau =\tau _0)}+\cdots  \label{1.21}
\end{eqnarray}

\noindent %
could be called {\it covariant Taylor row}.

A co-ordinate tensor basis $\partial _{A}\otimes dx^{B}$ at the point $%
x^{i}(\tau _{0}+d\tau )$ could be expressed by means of the tensor basis $%
\partial _{A}\otimes dx^{B}$ at the point $x^{i}(\tau _{0})$ of the curve $%
x^{i}(\tau )$ as 
\[
(\partial _{A}\otimes dx^{B})_{(\tau =\tau _{0}+d\tau )}=\left[ (\exp [d\tau
\cdot \frac{D}{d\tau }])(\partial _{A}\otimes dx^{B})\right] _{(\tau =\tau
_{0})}\text{ \ .} 
\]

Now, we can determine the covariant derivative $\nabla _uV$ of a tensor
field $V\in \otimes ^k\,_l(M)$ along a curve $x^i(\tau )$ with tangential
vector $u=d/d\tau $ at a given point $x^i(\tau =\tau _0)$ as 
\begin{equation}
\nabla _uV_{(\tau _0)}=\left( \frac{DV}{d\tau }\right) _{(\tau
_0)}=\lim_{d\tau \rightarrow 0}\frac{V_{(\tau _0+d\tau )}-V_{(\tau _0)}}{%
d\tau }\text{ , \ \ \ \ \ \ \ }d\tau =\varepsilon \ll 1\text{ \ .}
\label{1.22}
\end{equation}

By the use of the covariant derivative along a curve and the exponent of the
covariant differential operator we can find the geometrical interpretation
of the curvature tensor as well as the geometrical interpretation of the
torsion vector and the torsion tensor.

\subsection{Hodge (star) operator in spaces with affine connections and
metrics}

In many problems of theoretical physics the full anti-symmetric covariant
tensor fields (differential forms) are used. The transition from one
differential form of rank $p\leq n$ to a differential form of rank $n-p$,
where $n$ is the dimension of the differentiable manifold $M$ over which
differential forms are defined, is related to the existence of a map called 
{\it Hodge or star operator} \cite{Dubrovin}, \cite{von Westenholz}. Its
explicit action is given in $E_n$ (Euclidean spaces) and $V_n$- spaces, as
well as in spaces with one affine connection and metrics. Its generalization
for $(\overline{L}_n,g)$-spaces requires some additional considerations.
Usually the Hodge (star) operator is constructed by means of the permutation
(Levi-Civita) symbols. It maps a full covariant anti-symmetric tensor of
rank $(0,p)\equiv \,^a\otimes _p(M)\equiv \Lambda ^p(M)$ in a full covariant
anti-symmetric tensor of rank $(0,n-p)\equiv \,^a\otimes _{n-p}(M)\equiv
\Lambda ^{n-p}(M)$, where $\dim M=n$.

\subsubsection{Definition of the Hodge (star) operator}

Let $_aA:=A_{[i_1\cdots i_k]}\cdot dx^{i_1}\wedge \cdots \wedge dx^{i_k}$ be
a full covariant anti-symmetric tensor field of rank $k$, i.e. $_aA\in
\Lambda ^k(M)$. Let $_a\overline{A}:=A^{[j_1\cdots j_k]}\cdot \partial
_{j_1}\wedge \cdots \wedge \partial _{j_k}$ be the corresponding full
contravariant anti-symmetric tensor field of rank $k$, i.e. $_a\overline{A}%
\in \,_a\otimes ^k(M)$. $_a\overline{A}$ is obtained by the use of $_aA$ and
the contravariant metric $\overline{g}=g^{kl}\cdot \partial _k.\partial _l$%
\begin{eqnarray}
A^{[j_1\cdots j_k]} &:&=g^{j_1\overline{i}_1}\cdot g^{j_2\overline{i}%
_2}\cdot ...\cdot g^{j_k\overline{i}_k}\cdot A_{[i_1\cdots i_k]}\text{ ,} 
\nonumber \\
_a\overline{A} &=&A^{[j_1\cdots j_k]}\cdot \partial _{j_1}\wedge \cdots
\wedge \partial _{j_k}=  \nonumber \\
&=&g^{j_1\overline{i}_1}\cdot g^{j_2\overline{i}_2}\cdot ...\cdot g^{j_k%
\overline{i}_k}\cdot A_{[i_1\cdots i_k]}\cdot \partial _{j_1}\wedge \cdots
\wedge \partial _{j_k}\text{ .}  \label{1.23}
\end{eqnarray}

Let $d\omega :=\sqrt{-d_g}\cdot \varepsilon _{i_1\cdots i_n\cdot
}dx^{i_1}\wedge \cdots \wedge dx^{i_n}$ be the invariant volume element in $%
M $ ($\dim M=n$), $d_g=\det (g_{ij})<0$. Let $d\omega $ acts on $_a\overline{%
A} $ as a mapping which maps $_a\overline{A}$ in a new tensor field $%
*\,(_aA) $%
\begin{eqnarray}
d\omega &:&\,_a\overline{A}\rightarrow (d\omega )(_a\overline{A}%
):=*(_{a\,}A)\in \Lambda ^{n-k}(M)\text{ \thinspace ,}  \nonumber \\
\ast (_{a\,}A) &=&(d\omega )(_a\overline{A}):=\frac 1{k!}\cdot \sqrt{-d_g}%
\cdot \varepsilon _{i_1\cdots i_{n-k}j_1\cdots j_k}\cdot A^{[\overline{j}%
_1\cdots \overline{j}_k]}\cdot dx^{i_1}\wedge \cdots  \nonumber \\
\cdots \wedge dx^{i_{n-k}} &=&  \nonumber \\
&=&\frac 1{k!}\cdot \sqrt{-d_g}\cdot \varepsilon _{i_1\cdots
i_{n-k}j_1\cdots j_k}\cdot g^{\overline{j}_1\overline{l}_1}\cdot g^{%
\overline{j}_2\overline{l}_2}\cdot \cdots \cdot g^{\overline{j}_k\overline{l}%
_k}\cdot A_{[l_1\cdots l_k]}\cdot dx^{i_1}\wedge \cdots  \nonumber \\
\cdots \wedge dx^{i_{n-k}} &=&  \nonumber \\
&=&*(_aA)_{[i_1\cdots i_{n-k}]}\cdot dx^{i_1}\wedge \cdots \wedge
dx^{i_{n-k}}\text{ \thinspace \thinspace ,}  \label{1.24}
\end{eqnarray}
\begin{equation}
\ast (_aA)_{[i_1\cdots i_{n-k}]}=\frac 1{k!}\cdot \sqrt{-d_g}\cdot
\varepsilon _{i_1\cdots i_{n-k}j_1\cdots j_k}\cdot g^{\overline{j}_1%
\overline{l}_1}\cdot g^{\overline{j}_2\overline{l}_2}\cdot \cdots \cdot g^{%
\overline{j}_k\overline{l}_k}\cdot A_{[l_1\cdots l_k]}\text{ .}  \label{1.25}
\end{equation}

Therefore, the Hodge (star) operator $*$ could be considered as defined by
means of two mappings: $*=d\omega \circ \overline{{\cal G}}$:

(a) The mapping $\overline{{\cal G}}:\,_aA\rightarrow \overline{{\cal G}}%
(_aA):=\,_a\overline{A}$, where $_aA\in \,^a\otimes _k(M)\equiv \Lambda
^k(M) $, $_a\overline{A}\in \,^a\otimes ^k(M)$. The operator $\overline{%
{\cal G}}$ is an operator, mapping a full covariant anti-symmetric tensor
field into a full contravariant anti-symmetric tensor field.

(b) The mapping $d\omega :\,_a\overline{A}\rightarrow (d\omega )(_a\overline{%
A})=(d\omega )(\overline{{\cal G}}(_aA))=(d\omega \circ \overline{{\cal G}}%
)(_aA):=*(_{a\,}A)$, where $_a\overline{A}\in \,^a\otimes ^k(M)$, $%
*(_{a\,}A)\in \,^a\otimes _{n-k}(M)\equiv \Lambda ^{n-k}(M)$. The mapping $%
d\omega $ is an invariant $n$-form in $M$ with $\dim M=n$, and at the same
time, it is the invariant volume element in $M$. The form $d\omega $ acts on 
$_a\overline{A}$ by means of the contraction operator $S$ over $M$ applied $%
k $-times on $_a\overline{A}$.

\subsubsection{Properties of the Hodge (star) operator}

The main property of the star operator $*$ is related to its double action
on a full covariant anti-symmetric tensor field. Let us now calculate the
expression $*(*(_aA))$ by the use of the explicit definition of the star
operator $*$. 
\begin{eqnarray*}
\ast (*(_aA)) &=&(d\omega )(\overline{*(_aA)})\text{ ,} \\
\overline{\ast (_aA)} &=&\overline{*(_aA)}^{[i_1\cdots i_{n-k}]}\cdot
\partial _{i_1}\wedge \cdots \wedge \partial _{i_{n-k}}\text{ ,} \\
\overline{\ast (_aA)}^{[i_1\cdots i_{n-k}]} &=&g^{i_1\overline{m}_1}\cdot
\cdots \cdot g^{i_{n-k}\overline{m}_{n-k}}\cdot (*(_aA))_{[m_1\cdots
m_{n-k}]}=
\end{eqnarray*}
\begin{eqnarray}
&=&\frac 1{k!}\cdot \sqrt{-d_g}\cdot \varepsilon _{m_1\cdots
m_{n-k}j_1\cdots j_k}\cdot g^{i_1\overline{m}_1}\cdot \cdots \cdot g^{i_{n-k}%
\overline{m}_{n-k}}.g^{\overline{j}_1\overline{l}_1}\cdot g^{\overline{j}_2%
\overline{l}_2}\cdot \cdots \cdot  \nonumber \\
&&\cdot \cdots g^{\overline{j}_k\overline{l}_k}\cdot A_{[l_1\cdots l_k]}%
\text{ .}  \label{1.26}
\end{eqnarray}

Therefore, 
\[
\ast (*(_aA))=(d\omega )(\overline{*(_aA)})= 
\]
\begin{eqnarray*}
&=&\frac 1{(n-k)!}\cdot \sqrt{-d_g}\cdot \varepsilon _{i_1\cdots
i_kp_1\cdots p_{n-k}}(\overline{*(_aA)})^{[p_1\cdots p_{n-k}]}\cdot
dx^{i_1}\wedge \cdots \wedge dx^{i_k}= \\
&=&\frac 1{(n-k)!}\cdot \sqrt{-d_g}\cdot \varepsilon _{i_1\cdots
i_kp_1\cdots p_{n-k}}\cdot \\
&&\cdot \frac 1{k!}\cdot \sqrt{-d_g}\cdot \varepsilon _{m_1\cdots
m_{n-k}j_1\cdots j_k}\cdot g^{\overline{p}_1\overline{m}_1}\cdot \cdots
\cdot g^{\overline{p}_{n-k}\overline{m}_{n-k}}.g^{\overline{j}_1\overline{l}%
_1}\cdot g^{\overline{j}_2\overline{l}_2}\cdot \cdots \cdot \\
&&\cdot \cdots g^{\overline{j}_k\overline{l}_k}\cdot A_{[l_1\cdots l_k]}\cdot
\end{eqnarray*}
\[
\cdot dx^{i_1}\wedge \cdots \wedge dx^{i_k}= 
\]
\begin{eqnarray}
&=&\frac 1{(n-k)!}\cdot \frac 1{k!}\cdot (-d_g)\cdot \varepsilon _{i_1\cdots
i_kp_1\cdots p_{n-k}}\cdot  \nonumber \\
&&\cdot \varepsilon _{m_1\cdots m_{n-k}j_1\cdots j_k}\cdot g^{\overline{m}_1%
\overline{p}_1}\cdot \cdots \cdot g^{\overline{m}_{n-k}\overline{p}%
_{n-k}}.g^{\overline{j}_1\overline{l}_1}\cdot g^{\overline{j}_2\overline{l}%
_2}\cdot \cdots \cdot g^{\overline{j}_k\overline{l}_k}\cdot A_{[l_1\cdots
l_k]}\cdot  \nonumber \\
&&\cdot dx^{i_1}\wedge \cdots \wedge dx^{i_k}  \label{1.27}
\end{eqnarray}

Since 
\[
\varepsilon _{m_1\cdots m_{n-k}j_1\cdots j_k}\cdot g^{\overline{m}_1%
\overline{p}_1}\cdot \cdots \cdot g^{\overline{m}_{n-k}\overline{p}%
_{n-k}}.g^{\overline{j}_1\overline{l}_1}\cdot g^{\overline{j}_2\overline{l}%
_2}\cdot \cdots \cdot g^{\overline{j}_k\overline{l}_k}= 
\]
\begin{equation}
=\det (g^{\overline{i}\overline{j}})\cdot \varepsilon ^{p_1\cdots
p_{n-k}l_1\cdots l_k}\text{ ,}  \label{1.27a}
\end{equation}

\noindent we have 
\begin{eqnarray}
\ast (*(_aA)) &=&\frac 1{(n-k)!}\cdot \frac 1{k!}\cdot (-d_g)\cdot \det (g^{%
\overline{i}\overline{j}})\cdot \varepsilon _{i_1\cdots i_kp_1\cdots
p_{n-k}}\cdot \varepsilon ^{p_1\cdots p_{n-k}l_1\cdots l_k}\cdot
A_{[l_1\cdots l_k]}\cdot  \nonumber \\
&&\cdot dx^{i_1}\wedge \cdots \wedge dx^{i_k}\text{ .}  \label{1.29}
\end{eqnarray}

On the other side, 
\begin{eqnarray}
(-d_g)\cdot \det (g^{\overline{i}\overline{j}}) &=&-\det (g_{kl})\cdot (\det
g^{\overline{i}\overline{j}})=-\det (g_k^i)=-1\text{ ,}  \nonumber \\
\varepsilon _{i_1\cdots i_kp_1\cdots p_{n-k}}\cdot \varepsilon ^{p_1\cdots
p_{n-k}l_1\cdots l_k} &=&(-1)^{k\cdot (n-k)}\cdot (n-k)!\cdot \varepsilon
_{i_1\cdots i_k}\cdot \varepsilon ^{l_1\cdots l_k}=  \nonumber \\
&=&(-1)^{k\cdot (n-k)}\cdot (n-k)!\cdot a_{i_1\cdots i_k}^{l_1\cdots l_k}%
\text{ .}  \label{1.30}
\end{eqnarray}

Then, it follows for $*(*(_aA))$%
\begin{eqnarray}
\ast (*(_aA)) &=&-\,\frac 1{k!\cdot (n-k)!}\cdot (-1)^{k\cdot (n-k)}\cdot
(n-k)!\cdot a_{i_1\cdots i_k}^{l_1\cdots l_k}\cdot A_{[l_1\cdots l_k]}\cdot
dx^{i_1}\wedge \cdots \wedge dx^{i_k}=  \nonumber \\
&=&-\,\frac 1{k!}\cdot (-1)^{k\cdot (n-k)}\cdot a_{i_1\cdots i_k}^{l_1\cdots
l_k}\cdot A_{[l_1\cdots l_k]}\cdot dx^{i_1}\wedge \cdots \wedge
dx^{i_k}\,\,\,\,\text{.}  \label{1.31}
\end{eqnarray}

Since 
\begin{equation}
a_{i_1\cdots i_k}^{l_1\cdots l_k}\cdot A_{[l_1\cdots l_k]}=\frac{n!}{(n-k)!}%
\cdot A_{[i_1\cdots i_k]}\text{ ,}  \label{1.32}
\end{equation}

\noindent we have as a result for $*(*(_aA))$%
\begin{eqnarray}
\ast (*(_aA)) &=&-\frac{n!}{(n-k)!k!}\cdot (-1)^{k\cdot (n-k)}\cdot
A_{[i_1\cdots i_k]}\cdot dx^{i_1}\wedge \cdots \wedge dx^{i_k}=  \nonumber \\
&=&-\frac{n!}{(n-k)!k!}\cdot (-1)^{k\cdot (n-k)}\cdot \,_aA\text{ .}
\label{1.33}
\end{eqnarray}

If $\det (g_{ij})=d_g>0$ we have 
\begin{eqnarray}
\ast (*(_aA)) &=&\frac{n!}{(n-k)!k!}\cdot (-1)^{k\cdot (n-k)}\cdot
A_{[i_1\cdots i_k]}\cdot dx^{i_1}\wedge \cdots \wedge dx^{i_k}=  \nonumber \\
&=&\frac{n!}{(n-k)!k!}\cdot (-1)^{k\cdot (n-k)}\cdot \,_aA\text{ .}
\label{1.34}
\end{eqnarray}

Therefore, 
\begin{eqnarray}
\ast (*(_aA)) &=&\varepsilon \cdot \frac{n!}{(n-k)!k!}\cdot (-1)^{k\cdot
(n-k)}\cdot \,_aA=(*\circ *)(_aA)\text{ ,}  \nonumber \\
\ast \circ * &=&\text{ }\varepsilon \cdot \frac{n!}{(n-k)!k!}\cdot
(-1)^{k\cdot (n-k)}\cdot id\text{ ,}  \label{1.35}
\end{eqnarray}

\noindent with $\varepsilon =-1$ for $d_g<0$ and $\varepsilon =1$ for $d_g>0$%
. The operator $id$ is the identity operator. Since 
\begin{equation}
\frac{n!}{k!}=\frac{k!(k+1)!}{k!}=(k+1)!\text{ ,\thinspace \thinspace
\thinspace \thinspace \thinspace \thinspace \thinspace \thinspace }%
(n-k)!=(n-1)\cdot (n-2)\cdot \cdots \cdot (n-k)\text{ ,}  \label{1.36}
\end{equation}

\noindent it follows that 
\begin{equation}
\frac{n!}{(n-k)!k!}=\binom nk =C_n^k\text{ ,}  \label{1.37}
\end{equation}

\noindent where $C_n^k$ are binomial coefficients. Now, we can write as a
final result for $*\circ *$ 
\begin{eqnarray}
\ast \circ * &=&\text{ }\varepsilon \cdot \frac{n!}{(n-k)!k!}\cdot
(-1)^{k\cdot (n-k)}\cdot id=*\circ *=\text{ }\varepsilon \cdot C_n^k\cdot
(-1)^{k\cdot (n-k)}\cdot id=  \nonumber \\
&=&*\circ *=\text{ }\varepsilon \cdot \binom nk \cdot (-1)^{k\cdot
(n-k)}\cdot id\text{ .}  \label{1.38}
\end{eqnarray}

The Hodge (star) operator is a necessary tool if we further wish to find the
relation between the vortex (rotation) velocity tensor and the vortex
(rotation) vector.

\subsection{Covariant divergency of a mixed tensor field}

The operation of the covariant differentiation along a contravariant vector
field can be extended to covariant differentiation along a contravariant
tensor field.

The Lie derivative $\pounds _\xi u$ of a contravariant vector field $u$
along a contravariant vector field $\xi $ can be expressed by the use of the
covariant differential operators $\nabla _\xi $ and $\nabla _u$ in the form 
\[
\pounds _\xi u=\nabla _\xi u-\nabla _u\xi -T(\xi ,u)\text{ ,\thinspace
\thinspace \thinspace \thinspace \thinspace \thinspace \thinspace \thinspace
\thinspace \thinspace \thinspace \thinspace \thinspace \thinspace \thinspace
\thinspace \thinspace \thinspace \thinspace \thinspace \thinspace }\xi \text{%
, }u\in T(M)\text{ ,} 
\]

\noindent where $T(\xi ,u)$ is the contravariant torsion vector field 
\[
T(\xi ,u)=T_{\alpha \beta }\,^\gamma \cdot \xi ^\alpha \cdot u^\beta \cdot
e_\gamma =T_{ij}\,^k\cdot \xi ^i\cdot u^j\cdot \partial _k\text{ ,} 
\]

\noindent constructed by means of the components $T_{\alpha \beta }\,^\gamma 
$ (or $T_{ij}\,^k)$ of the contravariant torsion tensor field $T$.

The Lie derivative $\pounds _\xi V$ of a contravariant tensor field $%
V=V^A\cdot e_A=V^A\cdot \partial _A\in \otimes ^m(M)$ along a contravariant
vector field $\xi $ can be written on the analogy of the relation for $%
\pounds _\xi u$ and by the use of the covariant differential operator $%
\nabla _\xi $ and an operator $\nabla _V$ in the form \cite{Manoff-11} 
\[
\pounds _\xi V=\nabla _\xi V-\nabla _V\xi -T(\xi ,V)\text{ ,\thinspace
\thinspace \thinspace \thinspace \thinspace }\xi \in T(M)\,\,\text{,
\thinspace }V\in \otimes ^m(M)\text{ ,\thinspace \thinspace \thinspace
\thinspace } 
\]

\noindent where 
\[
\begin{array}{c}
\nabla _V\xi =-\xi ^\alpha \,_{/\beta }\cdot S_{B\alpha }\,^{A\beta }\cdot
V^B\cdot e_A=-\xi ^i\,_{;j}\cdot S_{Ci}\,^{Aj}\cdot V^C\cdot \partial _A 
\text{ ,} \\ 
T(\xi ,V)=T_{B\gamma }\,^A\cdot \xi ^\gamma \cdot V^B\cdot
e_A=T_{Ck}\,^A\cdot \xi ^k\cdot V^C\cdot \partial _A \text{ ,} \\ 
T_{B\gamma }\,^A=T_{\beta \gamma }\,^\alpha \cdot S_{B\alpha }\,^{A\beta }%
\text{ ,\thinspace \thinspace \thinspace \thinspace \thinspace \thinspace }%
T_{Ck}\,^A=T_{jk}\,^i\cdot S_{Ci}\,^{Aj}\text{ .}
\end{array}
\]

$\nabla _V\xi $ appears as a definition of the action of the operator $%
\nabla _V$ on the vector field $\xi $. Let we now consider more closely this
operator and its properties.

Let a mixed tensor field $K\in \otimes ^k\,_l(M)$ be given in a
non-co-ordinate (or co-ordinate) basis 
\[
\begin{array}{c}
K=K^C\,_D\cdot e_C\otimes e^D=K^{C_1\alpha }\,_D\cdot e_{C_1}\otimes
e_\alpha \otimes e^D\text{ ,} \\ 
e_C=e_{C_1}\otimes e_\alpha \text{ ,\thinspace \thinspace \thinspace
\thinspace \thinspace \thinspace \thinspace \thinspace \thinspace \thinspace
\thinspace \thinspace \thinspace \thinspace \thinspace \thinspace \thinspace
\thinspace \thinspace \thinspace }e^D=e^{\alpha _1}\otimes \cdots \otimes
e^{\alpha _l}\text{ .}
\end{array}
\]

The action of the operator $\nabla _V$ on the mixed tensor field $K$ can be
defined on the analogy of the action of $\nabla _V$ on a contravariant
vector field $\xi $%
\begin{equation}  \label{X.1.-1}
\nabla _VK=-K^{C_1\alpha }\,_{D/\beta }\cdot S_{B\alpha }\,^{A\beta }\cdot
V^B\cdot e_{C_1}\otimes e^D\otimes e_A\text{ ,}
\end{equation}

\noindent where $\nabla _V$ is the covariant differential operator along a
contravariant tensor field $V$%
\[
\nabla _V:K\Rightarrow \nabla _VK\text{ , \thinspace \thinspace \thinspace
\thinspace \thinspace \thinspace \thinspace \thinspace }K\in \otimes
^k\,_l(M)\text{ ,\thinspace \thinspace \thinspace \thinspace \thinspace
\thinspace }V\in \otimes ^m(M)\text{ ,\thinspace \thinspace \thinspace
\thinspace }\nabla _VK\in \otimes ^{k-1+m}\,_l(M)\text{ .} 
\]

{\it Remark}. There is an other possibility for a generalization of the
action of $\nabla _V$ on a mixed tensor field 
\[
\begin{array}{c}
\nabla _VK=-\sum_{m=1}^kK^{\alpha _1...\alpha _m...\alpha _k}\,_{D/\beta
}\cdot S_{B\alpha _m}\,^{A\beta }\cdot V^B\cdot e_{\alpha _1}\otimes \cdots
\otimes e_{\alpha _{m-1}}\otimes e_{\alpha _{m+1}}\otimes \\ 
\cdots \otimes e_{\alpha _k}\otimes e^D\otimes e_A\text{ .}
\end{array}
\]

The result of the action of this operator on a contravariant vector field $%
\xi $ is identical with the action of the above defined operator $\nabla _V$.

{\it Remark}. A covariant differential operator along a contravariant tensor
field can also be defined through its action on mixed tensor fields in the
form 
\[
\overline{\nabla }_VK=K^C\,_{D/\beta }\cdot V^{A_1\beta }\cdot e_C\otimes
e^D\otimes e_{A_1}\text{ .} 
\]

The operator $\overline{\nabla }_V$ differs from $\nabla _V$ in its action
on a contravariant vector field and does not appear as a generalization of
the already defined operator by its action on a contravariant vector field.
It appears as a new differential operator acting on mixed tensor fields.

The {\it covariant differential operator} $\nabla _V$ has the properties:

(a) Linear operator 
\[
\begin{array}{c}
\nabla _V(\alpha \cdot K_1+\beta \cdot K_2)=\alpha \cdot \nabla _VK_1+\beta
\cdot \nabla _VK_2\text{ ,} \\ 
\alpha \text{ , }\beta \in {\bf R}\text{ (or }{\bf C}\text{), \thinspace
\thinspace \thinspace \thinspace \thinspace \thinspace \thinspace \thinspace
\thinspace \thinspace \thinspace \thinspace \thinspace }K_1,K_2\in \otimes
^k\,_l(M)\text{ .}
\end{array}
\]

The proof of this property follows immediately from (\ref{X.1.-1}) and the
linear property of the covariant differential operator along a basic
contravariant vector field.

(b) Differential operator (not obeying the Leibniz rule) 
\[
\begin{array}{c}
\nabla _V(K\otimes S)=\nabla _{e_\beta }K\otimes \overline{S}\,^\beta
+K\otimes \nabla _VS\text{ ,} \\ 
K=K^A\,_B\cdot e_A\otimes e^B \text{ ,\thinspace \thinspace \thinspace
\thinspace \thinspace \thinspace }\nabla _{e_\beta }K=K^A\,_{B/\beta }\cdot
e_A\otimes e^B\text{ ,} \\ 
S= \widetilde{S}\,^C\,_D\cdot e_C\otimes e^D=\widetilde{S}\,^{C_1\alpha
}\,_D\cdot e_{C_1}\otimes e_\alpha \otimes e^D\text{ ,} \\ 
\overline{S}\,^\beta =-\widetilde{S}\,^{C_1\alpha }\,_D\cdot S_{E\alpha
}\,^{F\beta }\cdot V^E\cdot e_{C_1}\otimes e^D\otimes e_F\text{ .}
\end{array}
\]

The proof of this property follows from the action of the defined in (\ref
{X.1.-1}) operator $\nabla _V$ and the properties of the covariant
derivative of the product of the components of the tensor fields $K$ and $S$.

If the tensor field $V$ is given as a contravariant metric tensor field $%
\overline{g}$, then the covariant differential operator $\nabla _V$ ($V=%
\overline{g}$) will have additional properties connected with the properties
of the contravariant metric tensor field.

\begin{definition}
{\it Contravariant metric differential operator} $\nabla _{\overline{g}}$.
Covariant differential operator $\nabla _V$ for $V=\overline{g}$.
\end{definition}
%

By means of the relations 
\begin{equation}
\begin{array}{c}
-S_{B\alpha }\,^{A\beta }\cdot g^B\cdot e_A=(g_\alpha ^\sigma \cdot g^{\beta
\kappa }+g_\alpha ^\kappa \cdot g^{\beta \sigma })\cdot e_\sigma \otimes
e_\kappa = \\ 
=(g_\alpha ^\sigma \cdot g^{\beta \kappa }+g_\alpha ^\kappa \cdot g^{\beta
\sigma })\cdot e_\sigma .e_\kappa \text{ ,}
\end{array}
\label{X.1.-2}
\end{equation}
\[
e_\sigma .e_\kappa =\frac 12\cdot (e_\sigma \otimes e_\kappa +e_\kappa
\otimes e_\sigma )\text{ ,} 
\]
\begin{equation}
K^{C_1\alpha }\,_{D/\beta }\cdot g_\alpha ^\sigma =K^{C_1\sigma }\,_{D/\beta
}\text{ ,}  \label{X.1.-3}
\end{equation}

\noindent the action of the contravariant metric differential operator on a
mixed tensor field $K$ can be represented in the form 
\begin{equation}  \label{X.1.-4}
\begin{array}{c}
\nabla _{\overline{g}}K=(K^{C_1\sigma }\,_{D/\beta }\cdot g^{\beta \kappa
}+K^{C_1\kappa }\,_{D/\beta }\cdot g^{\beta \sigma })\cdot e_{C_1}\otimes
e^D\otimes e_\sigma \otimes e_\kappa = \\ 
(K^{C_1\sigma }\,_{D/\beta }\cdot g^{\beta \kappa }+K^{C_1\kappa
}\,_{D/\beta }\cdot g^{\beta \sigma }).e_{C_1}\otimes e^D\otimes e_\sigma
.e_\kappa \text{ .}
\end{array}
\end{equation}

The properties of the operator $\nabla _{\overline{g}}$ are determined
additionally by the properties of the contravariant metric tensor field of
second rank:

(a) $\nabla _{\overline{g}}:K\Rightarrow \nabla _{\overline{g}}K$
,\thinspace \thinspace \thinspace \thinspace $K\in \otimes ^k\,_l(M)$, $%
\nabla _{\overline{g}}K\in \otimes ^{k+1}\,_l(M)$.

(b) Linear operator 
\[
\nabla _{\overline{g}}(\alpha \cdot K_1+\beta \cdot K_2)=\alpha \cdot \nabla
_{\overline{g}}K_1+\beta \cdot \nabla _{\overline{g}}K_2\text{ . } 
\]

(c) Differential operator (not obeying the Leibniz rule) 
\begin{equation}  \label{X.1.-5}
\begin{array}{c}
\nabla _{\overline{g}}(K\otimes S)=\nabla _{e_\beta }K\otimes \overline{S}%
\,^\beta +K\otimes \nabla _{\overline{g}}S\text{ ,} \\ 
K\in \otimes ^k\,_l(M) \text{ ,\thinspace \thinspace \thinspace \thinspace
\thinspace \thinspace \thinspace \thinspace \thinspace \thinspace \thinspace 
}S\in \otimes ^m\,_r(M)\text{ ,} \\ 
K=K^A\,_B\cdot e_A\otimes e^B \text{ ,\thinspace \thinspace \thinspace
\thinspace \thinspace \thinspace }\nabla _{e_\beta }K=K^A\,_{B/\beta }\cdot
e_A\otimes e^B\text{ ,} \\ 
S= \widetilde{S}\,^C\,_D\cdot e_C\otimes e^D=\widetilde{S}\,^{C_1\alpha
}\,_D\cdot e_{C_1}\otimes e_\alpha \otimes e^D\text{ ,} \\ 
\overline{S}\,^\beta =(\widetilde{S}\,^{C_1\sigma }\,_D\cdot g^{\beta \kappa
}+\widetilde{S}\,^{C_1\kappa }\,_D\cdot g^{\beta \sigma })\cdot
e_{C_1}\otimes e^D\otimes e_\sigma .e_\kappa \text{ .}
\end{array}
\end{equation}

{\it Remark}. 
%
The definition of $\nabla _{\overline{g}}$ in (\ref{X.1.-4}) differs from
the definition in \cite{Manoff-2}, where $\nabla _{\overline{g}}\equiv 
\overline{\nabla }_{\overline{g}}$, i. e. the contravariant metric
differential operator is defined in the last case as a special case of the
covariant differential operator $\overline{\nabla }_V$ for $V=\overline{g}$.

%

The notion of covariant divergency of a mixed tensor field has been used in $%
V_4$-spaces for the determination of conditions for the existence of local
conserved quantities and in identities of the type of the first covariant
Noether identity. Usually, the covariant divergency of a contravariant or
mixed tensor field has been given in co-ordinate or non-co-ordinate basis in
the form 
\begin{equation}  \label{X.1.-6}
\delta K=K^{A\beta }\,_{B/\beta }\cdot e_A\otimes e^B=K^{Ci}\,_{D;i}\cdot
\partial _C\otimes dx^D\text{ ,}
\end{equation}

\noindent where 
\begin{equation}  \label{X.1.-7}
K^{A\beta }\,_{B/\beta }=K^{A\beta }\,_{B/\gamma }\cdot g_\beta ^\gamma 
\text{ ,\thinspace \thinspace \thinspace \thinspace \thinspace \thinspace
\thinspace \thinspace \thinspace }K^{Ci}\,_{D;i}=K^{Ci}\,_{D;k}\cdot g_i^k%
\text{ .\thinspace }
\end{equation}

For full anti-symmetric covariant tensor fields (differential forms) the
covariant divergency (called also codifferential) $\delta $ is defined by
means of the Hodge operator $*$, its reverse operator $*^{-1}$ and the
external differential operator $_a\overline{d}$ in the form \cite{Dubrovin}, 
\cite{von Westenholz} (pp. 147-149) 
\begin{equation}  \label{X.1.-8}
\delta =*^{-1}\circ \,_a\overline{d}\circ *\text{ .}
\end{equation}

{\it Remark}. 
%
%
The {\it Hodge operator} is constructed by means of the permutation
(Levi-Chivita) symbols. It maps a full covariant anti-symmetric tensor of
rank $(0,p)\equiv \,^a\otimes _pM)\equiv \Lambda ^p(M)$ in a full covariant
anti-symmetric tensor of rank $(0,n-p)\equiv \,^a\otimes _{n\,-\,p}(M)\equiv
\Lambda ^{n\,-\,p}(M)$, where $\dim M=n$, 
\[
\ast \,:\,_aA\rightarrow *\,_aA\text{ , \thinspace \thinspace \thinspace
\thinspace \thinspace \thinspace \thinspace \thinspace }_aA\in \Lambda ^p(M)\text{ , \thinspace \thinspace \thinspace \thinspace \thinspace \thinspace
\thinspace }*\,_aA\in \Lambda ^{n\,-\,p}(M)\text{ ,} 
\]

with 
\[
_aA=A_{[i_1...i_p]}.dx^{i_1}\wedge ...\wedge dx^{i_p}\text{ ,\thinspace
\thinspace \thinspace \thinspace \thinspace \thinspace \thinspace \thinspace
\thinspace \thinspace }*\,_aA=*A_{[j_1...j_{n\,-\,p}]}.dx^{j_1}\wedge
...\wedge dx^{j_{n\,-\,p}}\text{ ,} 
\]
\[
\ast A_{[j_1...j_{n\,-\,p}]}=\frac 1{p!}.\varepsilon
_{i_1...i_pj_1...j_{n\,-\,p}}.A^{[i_1...i_p]}\text{ , \thinspace \thinspace }A^{[i_1...i_p]}=g^{i_1\overline{k}_1}....g^{i_p\overline{k}_p}.A_{[k_1...k_p]}\text{ ,\thinspace \thinspace \thinspace \thinspace } 
\]
\[
\ast ^{-1}=(-1)^{p.(n-p)}.*\text{ .} 
\]

%

By the use of the contravariant metric differential operator $\nabla _{%
\overline{g}}$, the covariant metric tensor field $g$ and the contraction
operator one can introduce the notion of covariant divergency of a mixed
tensor field $K$ with finite rank.

\begin{definition}
{\bf \ } {\it Covariant divergency }$\delta K${\it \ of a mixed tensor field}
$K$\[
\delta K=\frac 12.[\nabla _{\overline{g}}K]g=K^{A\beta }\,_{B/\beta
}.e_A\otimes e^B=K^{Ci}\,_{D;i}.\partial _C\otimes dx^D\text{ ,} 
\]
\end{definition}
%

\noindent where 
\[
\begin{array}{c}
K=K^{A\beta }\,_B\cdot e_A\otimes e_\beta \otimes e^B=K^{Ci}\,_D\cdot
\partial _C\otimes \partial _i\otimes dx^D \text{ ,} \\ 
K\in \otimes ^k\,_l(M)\text{ , \thinspace \thinspace \thinspace \thinspace
\thinspace }k\geq 1\text{ .}
\end{array}
\]

$\delta $ is called operator of the covariant divergency 
\[
\delta :K\Rightarrow \delta K\text{ ,\thinspace \thinspace \thinspace
\thinspace \thinspace \thinspace \thinspace \thinspace \thinspace }K\in
\otimes ^k\,_l(M)\text{ ,\thinspace \thinspace \thinspace \thinspace
\thinspace \thinspace \thinspace \thinspace \thinspace \thinspace }\delta
K\in \otimes ^{k-1}\,_l(M)\text{ ,\thinspace \thinspace \thinspace
\thinspace }k\geq 1\text{ .} 
\]

{\it Remark}. 
The symbol $\delta $ has also been introduced for the variation operator.
Both operators are different from each other and can easily be
distinguished. Ambiguity would occur only if the symbol $\delta $ is used
out of the context. In such a case, the definition of the symbol $\delta $
is necessary.
%

The properties of the operator of the covariant divergency $\delta $ are
determined by the properties of the contravariant metric differential
operator, the contraction operator and the metric tensor fields $g$ and $%
\overline{g}$

(a) The operator of the covariant divergency $\delta $ is a linear operator 
\begin{equation}
\begin{array}{c}
\delta (\alpha \cdot K_1+\beta \cdot K_2)=\alpha \cdot \delta K_1+\beta
\cdot \delta K_2\text{ ,\thinspace \thinspace } \\ 
\text{\thinspace }\alpha ,\beta \in {\bf R}\text{ (or }{\bf C}\text{)
,\thinspace \thinspace \thinspace \thinspace \thinspace \thinspace
\thinspace \thinspace \thinspace \thinspace \thinspace \thinspace \thinspace
\thinspace \thinspace \thinspace \thinspace }K_1,K_2\in \otimes ^k\,_l(M)%
\text{ .}
\end{array}
\label{X.1.-9}
\end{equation}

The proof of this property follows immediately from the definition of the
covariant divergency.

(b) Action on a tensor product of tensor fields 
\begin{equation}  \label{X.1.-10}
\delta (K\otimes S)=\overline{\nabla }_SK+K\otimes \delta S\text{ ,}
\end{equation}

\noindent where 
\begin{equation}  \label{X.1.-11}
\begin{array}{c}
K=K^A\,_B\cdot e_A\otimes e^B \text{ ,\thinspace \thinspace \thinspace
\thinspace \thinspace \thinspace }S=S^{C\beta }\,_D\cdot e_C\otimes e_\beta
\otimes e^D\text{ ,} \\ 
\overline{\nabla }_SK=K^A\,_{B/\beta }\cdot S^{C\beta }\,_D\cdot e_A\otimes
e^B\otimes e_C\otimes e^D\text{ \thinspace \thinspace \thinspace \thinspace
\thinspace \thinspace \thinspace (see above }\overline{\nabla }_V\text{) .}
\end{array}
\end{equation}

The proof of this property follows from the properties of $\nabla _{%
\overline{g}}$ and from the relations 
\begin{equation}  \label{X.1.-12}
\frac 12[\nabla _{e_\beta }K\otimes \overline{S}\,^\beta ]g=K^A\,_{B/\beta
}\cdot S^{C\beta }\,_D\cdot e_A\otimes e^B\otimes e_C\otimes e^D\text{ ,}
\end{equation}
\begin{equation}  \label{X.1.-13}
\frac 12[K\otimes \nabla _{\overline{g}}S]g=K\otimes \frac 12\cdot [\nabla _{%
\overline{g}}S]g=K\otimes \delta S\text{ .}
\end{equation}

(c) Action on a contravariant vector field $u$%
\begin{equation}  \label{X.1.-14}
\delta u=\frac 12\cdot [\nabla _{\overline{g}}u]g=u^\beta \,_{\,/\beta
}=u^i\,_{;i}\text{ .}
\end{equation}

(d) Action on the tensor product of two contravariant vector fields $u$ and $%
v$%
\begin{equation}  \label{X.1.-15}
\delta (u\otimes v)=\nabla _vu+\delta u\cdot v\text{ ,\thinspace \thinspace
\thinspace \thinspace \thinspace \thinspace \thinspace \thinspace }\overline{%
\nabla }_vu=\nabla _vu\text{ .}
\end{equation}

(e) Action of the product of an invariant function $L$ and a mixed tensor
field $K$%
\begin{equation}
\delta (L.K)=\overline{\nabla }_KL+L\cdot \delta K\text{ ,}  \label{X.1.-16}
\end{equation}
\begin{equation}
\begin{array}{c}
\overline{\nabla }_KL=L_{/\beta }\cdot K^{A\beta }\,_B\cdot e_A\otimes e^B%
\text{ ,\thinspace \thinspace \thinspace \thinspace \thinspace \thinspace }%
\delta K=K^{A\beta }\,_{B/\beta }\cdot e_A\otimes e^B\text{ ,} \\ 
L_{/\beta }=e_\beta L\text{ ,\thinspace \thinspace \thinspace \thinspace
\thinspace \thinspace \thinspace \thinspace \thinspace \thinspace \thinspace
\thinspace \thinspace \thinspace \thinspace \thinspace \thinspace }%
L_{;i}=L_{,i}\text{ ,} \\ 
K=K^{A\beta }\,_B\cdot e_A\otimes e_\beta \otimes e^B\in \otimes ^k\,_l(M)%
\text{ .}
\end{array}
\label{X.1.-17}
\end{equation}

{\it Special case}: Action of the product of an invariant function $L$ and
the contravariant metric tensor $\overline{g}$: 
\begin{equation}
\delta (L\cdot \overline{g})=(L_{/\beta }\cdot g^{\alpha \beta }+L\cdot
g^{\alpha \beta }\,_{/\beta })=(L_{,j}\cdot g^{ij}+L\cdot g^{ij}\,_{;j})%
\text{ .}  \label{X.1.-18}
\end{equation}

(f) Action on an anti-symmetric tensor product of two contravariant vector
fields $u$ and $v$%
\begin{equation}  \label{X.1.-19}
\begin{array}{c}
\delta (u\wedge v)=\frac 12\cdot (\nabla _vu-\nabla _uv+\delta v\cdot
u-\delta u\cdot v)= \\ 
=-\frac 12\cdot [\pounds _uv+T(u,v)+\delta u\cdot v-\delta v\cdot u]\text{ .}
\end{array}
\end{equation}

(g) Action on a full anti-symmetric contravariant tensor field $A$ of second
rank 
\begin{equation}  \label{X.1.-20}
\begin{array}{c}
\delta A=\frac 12\cdot (A^{\alpha \beta }-A^{\beta \alpha })_{/\beta }\cdot
e_\alpha =\frac 12\cdot (A^{ij}-A^{ji})_{;j}\cdot \partial _i \text{ ,} \\ 
A=A^{\alpha \beta }\cdot e_\alpha \wedge e_\beta =A^{ij}\cdot \partial
_i\wedge \partial _j\text{ ,\thinspace \thinspace \thinspace \thinspace
\thinspace \thinspace }A^{\alpha \beta }=-\,A^{\beta \alpha }\text{ .}
\end{array}
\end{equation}

(h) Action on a tensor product of a contravariant vector field $u$,
multiplied with an invariant function, and a covariant vector field $g(v)$
with the contravariant vector field $v$%
\begin{equation}  \label{X.1.-21}
\begin{array}{c}
\delta (\varepsilon \cdot u\otimes g(v))=(u\varepsilon )\cdot
g(v)+\varepsilon \cdot [\delta u\cdot g(v)+(\nabla _ug)(v)+g(\nabla _uv)]=
\\ 
=[u\varepsilon +\varepsilon \cdot \delta u]\cdot g(v)+\varepsilon \cdot
[(\nabla _ug)(v)+g(\nabla _uv)] \text{ ,} \\ 
\varepsilon \in C^r(M)\text{ , }\,\,\,\,\,\,\,\,\varepsilon ^{\prime
}(x^{k^{\prime }})=\varepsilon (x^k)\text{ ,\thinspace \thinspace \thinspace
\thinspace \thinspace \thinspace \thinspace \thinspace }u,v\in T(M)\text{ .}
\end{array}
\end{equation}

{\it Special case}: $v\equiv u$: 
\begin{equation}
\delta (\varepsilon \cdot u\otimes g(u))=[u\varepsilon +\varepsilon \cdot
\delta u]\cdot g(u)+\varepsilon \cdot [(\nabla _ug)(u)+g(a)]\text{
,\thinspace \thinspace \thinspace \thinspace \thinspace \thinspace
\thinspace \thinspace }\nabla _uu=a\text{ .}  \label{X.1.-22}
\end{equation}

{\it Special case}: $\varepsilon =1$: 
\begin{equation}  \label{X.1.-23}
\delta (u\otimes g(v))=\delta u\cdot g(v)+(\nabla _ug)(v)+g(\nabla _uv)\text{
.}
\end{equation}

\subsection{Covariant divergency of a mixed tensor field of second rank}

From the definition of the covariant divergency $\delta K$ of a mixed tensor
field $K$, the explicit form of the covariant divergency of tensor fields of
second rank of the type 1 or 2 follows as 
\begin{equation}
\delta G=\frac 12[\nabla _{\overline{g}}G]g=G_\alpha \,^\beta \,_{/\beta
}\cdot e^\alpha =G_i\,^j\,_{;j}\cdot dx^i\text{ ,}  \label{X.2.-1}
\end{equation}
\begin{equation}
\delta \overline{G}=\frac 12[\nabla _{\overline{g}}\overline{G}]g=\overline{G%
}\,^\beta \,_{\alpha /\beta }\cdot e^\alpha =\overline{G}\,^j\,_{i;j}\cdot
dx^i\text{ .}  \label{X.2.-2}
\end{equation}

By the use of the relations (\ref{X.1.-14}) $\div \,$(\ref{X.1.-19}), (\ref
{X.1.-21}) $\div \,$(\ref{X.1.-23}), and the expression [see (\ref{X.1.-10}) 
$\div $ (\ref{X.1.-13})] 
\begin{equation}  \label{X.2.-3}
\overline{\nabla }_v(g(u))=\nabla _v(g(u))=(\nabla _vg)(u)+g(\nabla _vu)%
\text{ ,}
\end{equation}
\begin{equation}  \label{X.2.-4}
\delta (g(u)\otimes v)=\delta v\cdot g(u)+(\nabla _vg)(u)+g(\nabla _vu)\text{
,}
\end{equation}
\begin{equation}  \label{X.2.-5}
\delta ((^GS)g)=(g_{\alpha \overline{\gamma }}\cdot \,^GS^{\beta \gamma
})_{/\beta }\cdot e^\alpha \text{ ,}
\end{equation}

\noindent the covariant divergency of the representation of $G$ by means the
rest mass density $\rho _G$ ($\varepsilon _G=\rho _G$) 
\[
G=\rho _G\cdot u\otimes g(u)+u\otimes g(^G\pi )+\,^Gs\otimes g(u)+(^GS)g, 
\]

\noindent can be found in the form ($\nabla _uu=a$) 
\begin{equation}  \label{X.2.-6}
\begin{array}{c}
\delta G=\rho _G\cdot g(a)+(u\rho _G+\rho _G\cdot \delta u+\delta ^Gs)\cdot
g(u)+\delta u\cdot g(^G\pi )+g(\nabla _u\,^G\pi )+ \\ 
+\,\,\,g(\nabla _{^Gs}u)+\rho _G\cdot (\nabla _ug)(u)+(\nabla _ug)(^G\pi
)+(\nabla _{^Gs}g)(u)+\delta ((^GS)g)\text{ .}
\end{array}
\end{equation}

$\overline{g}(\delta G)$ will have the form 
\begin{equation}  \label{X.2.-7}
\begin{array}{c}
\overline{g}(\delta G)=\rho _G\cdot a+(u\rho _G+\rho _G\cdot \delta u+\delta
^Gs)\cdot u+\delta u\cdot \,^G\pi +\nabla _u\,^G\pi + \\ 
+\,\,\,\nabla _{^Gs}u+\rho _G\cdot \overline{g}(\nabla _ug)(u)+\overline{g}%
(\nabla _ug)(^G\pi )+\overline{g}(\nabla _{^Gs}g)(u)+\overline{g}(\delta
((^GS)g))\text{ .}
\end{array}
\end{equation}

In a co-ordinate basis $\delta G$ and $\overline{g}(\delta G)$ will have the
forms 
\begin{equation}  \label{X.2.-10}
\begin{array}{c}
G_i\,^j\,_{;j}=\rho _G\cdot a_i+(\rho _{G,j}\cdot u^j+\rho _G\cdot
u^j\,_{;j}+\,^Gs^j \text{ }_{;j})\cdot u_i+u^j\,_{;j}\cdot \,^G\pi _i+ \\ 
+\;g_{i \overline{j}}\cdot (^G\pi ^j\,_{;k}\cdot u^k+u^j\,_{;k}\cdot
\,^Gs^k)+g_{ij;k}\cdot (\rho _G\cdot u^k\cdot u^{\overline{j}}+u^k\cdot
\,^G\pi ^{\overline{j}}+\,^Gs^k\cdot u^{\overline{j}})+ \\ 
+\,\,(g_{i\overline{k}}\cdot ^GS^{jk})_{;j}\text{ ,}
\end{array}
\end{equation}
\[
a_i=g_{i\overline{j}}\cdot a^j\text{ ,\thinspace \thinspace \thinspace
\thinspace \thinspace }a^i=u^i\,_{;j}\cdot u^j\text{ ,\thinspace \thinspace
\thinspace \thinspace }\rho _{G;i}=\rho _{G,i}\text{ ,\thinspace \thinspace
\thinspace \thinspace \thinspace \thinspace }u_i=g_{i\overline{k}}\cdot u^k%
\text{ ,\thinspace \thinspace \thinspace \thinspace }^G\pi _i=g_{i\overline{j%
}}\cdot ^G\pi ^j\text{ ,} 
\]
\begin{equation}  \label{X.2.-11}
\begin{array}{c}
g^{i \overline{k}}\cdot G_k\,^j\,_{;j}=\rho _G\cdot a^i+(\rho _{G,j}\cdot
u^j+\rho _G\cdot u^j\,_{;j}+\,^Gs^j\text{ }_{;j})\cdot u^i+u^j\,_{;j}\cdot
\,^G\pi ^i+ \\ 
+\;^G\pi ^i\,_{;j}\cdot u^j+u^i\,_{;j}\cdot \,^Gs^j+g^{i \overline{l}}\cdot
g_{lj;k}\cdot (\rho _G\cdot u^k\cdot u^{\overline{j}}+u^k\cdot \,^G\pi ^{%
\overline{j}}+\,^Gs^k\cdot u^{\overline{j}})+ \\ 
+\,\,g^{i\overline{l}}\cdot (g_{lk}\cdot \,^GS^{jk})_{;j}\text{ .}
\end{array}
\end{equation}

The relation between $\delta G$ and $\delta \overline{G}$ follows from the
relation between $G$ and $\overline{G}$%
\begin{equation}  \label{X.2.-18}
\overline{G}=g(G)\overline{g}:\delta \overline{G}=\delta (g(G)\overline{g}%
)=\frac 12[\nabla _{\overline{g}}(g(G)\overline{g})]\text{ .}
\end{equation}

The {\it covariant divergency of the Kronecker tensor field} can be found in
an analogous way as the covariant divergency of a tensor field of second
rank of the type 1, since $Kr=g_\beta ^\alpha \cdot e_\alpha \otimes e^\beta
=g_j^i\cdot \partial _i\otimes dx^j$%
\begin{equation}  \label{X.2.-37}
\delta Kr=\frac 12[\nabla _{\overline{g}}Kr]g=g_\alpha ^\beta \,_{/\beta
}\cdot e^\alpha =g_{i;j}^j\cdot dx^i\text{ .}
\end{equation}

If we use the representation of $Kr$%
\[
Kr=\frac 1e\cdot k\cdot u\otimes g(u)+u\otimes g(^{Kr}\pi )+\,^{Kr}s\otimes
g(u)+(^{Kr}S)g\text{ ,} 
\]

\noindent then the covariant divergency $\delta Kr$ can be written in the
form 
\begin{equation}  \label{X.2.-38}
\begin{array}{c}
\delta Kr=\frac 1e\cdot k\cdot g(a)+[u(\frac 1e\cdot k)+\frac 1e\cdot k\cdot
\delta u+\delta ^{Kr}s]\cdot g(u)+ \\ 
+\delta u\cdot g(^{Kr}\pi )+g(\nabla _u\,^{Kr}\pi )+g(\nabla _{^{Kr}s}u)+ \\ 
+\,\frac 1e\cdot k\cdot (\nabla _ug)(u)+(\nabla _ug)(^{Kr}\pi )+(\nabla
_{^{Kr}s}g)(u)+\delta ((^{Kr}S)g)\text{ ,}
\end{array}
\end{equation}

\noindent or in the form 
\begin{equation}  \label{X.2.-39}
\begin{array}{c}
\overline{g}(\delta Kr)=\frac 1e\cdot k\cdot a+[u(\frac 1e\cdot k)+\frac
1e\cdot k\cdot \delta u+\delta ^{Kr}s]\cdot u+ \\ 
+\,\delta u\cdot \,^{Kr}\pi +\nabla _u\,^{Kr}\pi +\nabla _{^{Kr}s}u+ \\ 
+\,\frac 1e\cdot k\cdot \overline{g}(\nabla _ug)(u)+\overline{g}(\nabla
_ug)(^{Kr}\pi )+\overline{g}(\nabla _{^{Kr}s}g)(u)+\overline{g}(\delta
((^{Kr}S)g))\text{ .}
\end{array}
\end{equation}

In a co-ordinate basis $\delta Kr$ and $\overline{g}(\delta Kr)$ will have
the forms 
\begin{equation}
\begin{array}{c}
g_i^j\,_{;j}=\frac 1e\cdot k\cdot a_i+[(\frac 1e\cdot k)_{,j}\cdot u^j+\frac
1e\cdot k\cdot u^j\,_{;j}+\,^{Kr}s^j\,_{;j}]\cdot u_i+ \\ 
+\,\,u^j\,_{;j}\cdot \,^{Kr}\pi _i+g_{i\overline{j}}\cdot (^{Kr}\pi
^j\,_{;k}\cdot u^k+u^j\,_{;k}\cdot \,^{Kr}s^k)+ \\ 
+\,g_{ij;k}\cdot (\frac 1e\cdot k\cdot u^{\overline{j}}\cdot u^k+\,^{Kr}\pi
^{\overline{j}}\cdot u^k+u^{\overline{j}}\cdot \,^{Kr}s^k)+(g_{i\overline{k}%
}\cdot \,^{Kr}S^{jk})_{;j}\text{ ,}
\end{array}
\label{X.2.-42}
\end{equation}
\begin{equation}
\begin{array}{c}
g^{i\overline{k}}\cdot g_k^j\,_{;j}=\frac 1e\cdot k\cdot a^i+[(\frac 1e\cdot
k)_{,j}\cdot u^j+\frac 1e\cdot k\cdot u^j\,_{;j}+\,^{Kr}s^j\,_{;j}]\cdot u^i+
\\ 
+\,\,u^j\,_{;j}\cdot \,^{Kr}\pi ^i+\,^{Kr}\pi ^i\,_{;j}\cdot
u^j+u^i\,_{;j}\cdot \,^{Kr}s^j+ \\ 
+\,g^{i\overline{l}}\cdot g_{lj;k}\cdot (\frac 1e\cdot k\cdot u^{\overline{j}%
}\cdot u^k+\,^{Kr}\pi ^{\overline{j}}\cdot u^k+u^{\overline{j}}\cdot
\,^{Kr}s^k)+g^{i\overline{l}}\cdot (g_{l\overline{k}}\cdot
\,^{Kr}S^{jk})_{;j}\text{ .}
\end{array}
\label{X.2.-43}
\end{equation}

\section{Geometrical interpretation of the curvature tensor}

The geometrical interpretation of the curvature tensor is well known for
spaces with one affine connection \cite{Raschewski} but all considerations,
related to this topic are made in a given co-ordinate basis and not in an
index-free manner. We will try now to find a reasonable geometrical
interpretation of the curvature tensor in a covariant (index-free) manner
using the introduced above mathematical tools.

Let us now consider a closed infinitesimal contour (quadrangle) $ACDBA$
build by a congruence of two parametric curves $x^i(\tau ,\lambda )$. Let
the tangent vectors to the curve $x^i(\tau ,\lambda =\lambda _0=$ const.$)$
and the curve $x^i(\tau =\tau _0=$ const.$,\,\lambda )$ be $u=d/d\tau $ and $%
d/d\lambda $ respectively.

\subsection{Covariant transport of a vector on two different paths from a
point to an other point of a closed infinitesimal contour}

Let us now consider a covariant transport of two vectors from one to an
other point of a closed infinitesimal contour $ACDBA$ on two different
paths: from point $A$ to point $D$ across the point $C$ and from point $A$
to point $D$ across the point $B$.

The point $A$ has the co-ordinates $x^i(\tau _0$, $\lambda _0)$. The vectors 
$u$ and $\xi $ at the point $A$ could be denoted as $u_{(\tau _0,\lambda
_0)} $ and $\xi _{(\tau _0,\lambda _0)}$.

Then the vectors $\overline{u}=u_{(\tau _0+d\tau ,\lambda _0)}$ and $%
\overline{\xi }=\xi _{(\tau _0+d\tau ,\lambda _0)}$ at the point $C$ [with
co-ordinates $x^i(\tau _0+d\tau ,\lambda _0)$] could be related to the
vectors $u_{(\tau _0,\lambda _0)}$ and $\xi _{(\tau _0,\lambda _0)}$ at the
point $A$ with co-ordinates $x^i(\tau _0,\lambda _0)$ by means of the
relations 
\begin{equation}
\stackunder{from\,\,p.\,A\,}{C}:\overline{u}=u_{(\tau _0+d\tau ,\lambda
_0)}=\left[ (\exp [d\tau \cdot \frac D{d\tau }])u\right] _{(\tau _0,\lambda
_0)}\text{ \thinspace \thinspace \thinspace \thinspace ,}  \label{1.39a}
\end{equation}
\begin{equation}
\stackunder{from\,\,p.\,A}{C}:\overline{\xi }=\xi _{(\tau _0+d\tau ,\lambda
_0)}=\left[ (\exp [d\tau \cdot \frac D{d\tau }])\xi \right] _{(\tau
_0,\lambda _0)}\text{\thinspace \thinspace \thinspace \thinspace \thinspace
\thinspace \thinspace .}  \label{1.39}
\end{equation}

The vectors $\widetilde{u}=u_{(\tau _0,\lambda _0+d\lambda )}$ and $%
\widetilde{\xi }=\xi _{(\tau _0,\lambda _0+d\lambda )}$ at point $B$ with
co-ordinates $x^i(\tau _0,\lambda _0+d\lambda )$ could be related to the
vectors $u_{(\tau _0,\lambda _0)}$ and $\xi _{(\tau _0,\lambda _0)}$ by
means of the relations 
\begin{equation}
\stackunder{from\,\,\,p.\,A}{B}:\widetilde{u}=u_{(\tau _0,\lambda
_0+d\lambda )}=\left[ (\exp [d\lambda \cdot \frac D{d\lambda }])u\right]
_{(\tau _0,\lambda _0)}\text{ \thinspace \thinspace \thinspace \thinspace ,}
\label{1.40}
\end{equation}
\begin{equation}
\stackunder{from\,\,\,p.\,A}{B}:\widetilde{\xi }=\xi _{(\tau _0,\lambda
_0+d\lambda )}=\left[ (\exp [d\lambda \cdot \frac D{d\lambda }])\xi \right]
_{(\tau _0,\lambda _0)}\text{\thinspace \thinspace \thinspace \thinspace
\thinspace \thinspace \thinspace .}  \label{1.41}
\end{equation}

The vectors $\widetilde{\widetilde{u}}=u_{(\tau _0+d\tau ,\lambda
_0+d\lambda )}$ and $\widetilde{\widetilde{\xi }}=\xi _{(\tau _0+d\tau
,\lambda _0+d\lambda )}$, obtained by the transport of the vectors $%
\widetilde{u}=u_{(\tau _0,\lambda _0+d\lambda )}$ and $\widetilde{\xi }=\xi
_{(\tau _0,\lambda _0+d\lambda )}$ along the curve $x^i(\tau ,\lambda
_0+d\lambda )$ from the point $B$ with co-ordinates $x^i(\tau _0,\lambda
_0+d\lambda )$ to the point $D$ with co-ordinates $x^i(\tau _0+d\tau
,\lambda _0+d\lambda )$, could be related to the vectors $\widetilde{u}%
=u_{(\tau _0,\lambda _0+d\lambda )}$ and $\widetilde{\xi }=\xi _{(\tau
_0,\lambda _0+d\lambda )}$ and then to the vectors $u_{(\tau _0,\lambda _0)}$
and $\xi _{(\tau _0,\lambda _0)}$ correspondingly by means of the relations 
\begin{eqnarray}
\stackunder{from\,\,\,p.\,B}{D} &:&\widetilde{\widetilde{u}}=u_{(\tau
_0+d\tau ,\lambda _0+d\lambda )}=\left[ (\exp [d\tau \cdot \frac D{d\tau
}])u\right] _{(\tau _0,\lambda _0+d\lambda )}=  \nonumber \\
&=&\left[ (\exp [d\tau \cdot \frac D{d\tau }])(\exp [d\lambda \cdot \frac
D{d\lambda }])u\right] _{(\tau _0,\lambda _0)}\,\,\,\,\,\,\,\,\text{,}
\label{1.42}
\end{eqnarray}
\begin{eqnarray}
\stackunder{from\,\,p.\,B}{D} &:&\widetilde{\widetilde{\xi }}=\xi _{(\tau
_0+d\tau ,\lambda _0+d\lambda )}=\left[ (\exp [d\tau \cdot \frac D{d\tau
}])\xi \right] _{(\tau _0,\lambda _0+d\lambda )}=  \nonumber \\
&=&\left[ (\exp [d\tau \cdot \frac D{d\tau }])(\exp [d\lambda \cdot \frac
D{d\lambda }])\xi \right] _{(\tau _0,\lambda _0)}\text{ \thinspace
\thinspace \thinspace \thinspace \thinspace \thinspace .}  \label{1.43}
\end{eqnarray}

Until now we have considered the transport of the vectors $u$ and $\xi $
from point $A$ to point $C$ and from the point $A$ to the point $D$ across
the point $B$.

On the other side, the vectors $\overline{\overline{u}}=u_{(\tau _0+d\tau
,\lambda _0+d\lambda )}$ and $\overline{\overline{\xi }}=\xi _{(\tau
_0+d\tau ,\lambda _0+d\lambda )}$, obtained by the transport of the vectors $%
\overline{u}=u_{(\tau _0+d\tau ,\lambda _0)}$ and $\overline{\xi }=\xi
_{(\tau _0+d\tau ,\lambda _0)}$ along the curve $x^i(\tau _0+d\tau ,\lambda
) $ from point $C$ with co-ordinates $x^i(\tau _0+d\tau ,\lambda _0)$ to the
point $D$ with co-ordinates $x^i(\tau _0+d\tau ,\lambda _0+d\lambda )$,
could be related to the vectors $\overline{u}=u_{(\tau _0+d\tau ,\lambda
_0)} $ and $\overline{\xi }=\xi _{(\tau _0+d\tau ,\lambda _0)}$ and then to
the vectors $u_{(\tau _0,\lambda _0)}$ and $\xi _{(\tau _0,\lambda _0)}$
correspondingly by means of the relations 
\begin{eqnarray}
\stackunder{from\,\,p.\,C}{D} &:&\overline{\overline{u}}=u_{(\tau _0+d\tau
,\lambda _0+d\lambda )}=\left[ (\exp [d\lambda \cdot \frac D{d\lambda
}])u\right] _{(\tau _0+d\tau ,\lambda _0)}=  \nonumber \\
&=&\left[ (\exp [d\lambda \cdot \frac D{d\lambda }])(\exp [d\tau \cdot \frac
D{d\tau }])u\right] _{(\tau _0,\lambda _0)}\text{ \thinspace \thinspace
\thinspace \thinspace \thinspace ,}  \label{1.44}
\end{eqnarray}
\begin{eqnarray}
\stackunder{from\,\,p.\,C}{D} &:&\overline{\overline{\xi }}=\xi _{(\tau
_0+d\tau ,\lambda _0+d\lambda )}=\left[ (\exp [d\lambda \cdot \frac
D{d\lambda }])\xi \right] _{(\tau _0+d\tau ,\lambda _0)}=  \nonumber \\
&=&\left[ (\exp [d\lambda \cdot \frac D{d\lambda }])(\exp [d\tau \cdot \frac
D{d\tau }])\xi \right] _{(\tau _0,\lambda _0)}\text{ \thinspace \thinspace
\thinspace \thinspace \thinspace \thinspace .}  \label{1.45}
\end{eqnarray}

For an other contravariant vector field $v$ we have analogous relations as
for the vector fields $u$ and $\xi $ if the vector $v$ is transported along
the infinitesimal contour $ACDBA$ from point $A$ to point $D$ across point $%
C $ or from point $A$ to point $D$ across point $B$

\begin{eqnarray}
\stackunder{from\,\,\,p.\,B}{D} &:&\widetilde{\widetilde{v}}=v_{(\tau
_0+d\tau ,\lambda _0+d\lambda )}=\left[ (\exp [d\tau \cdot \frac D{d\tau
}])v\right] _{(\tau _0,\lambda _0+d\lambda )}=  \nonumber \\
&=&\left[ (\exp [d\tau \cdot \frac D{d\tau }])(\exp [d\lambda \cdot \frac
D{d\lambda }])v\right] _{(\tau _0,\lambda _0)}\,\,\,\,\,\,\,\,\text{,}
\label{1.46}
\end{eqnarray}
\begin{eqnarray}
\stackunder{from\,\,p.\,C}{D} &:&\overline{\overline{v}}=v_{(\tau _0+d\tau
,\lambda _0+d\lambda )}=\left[ (\exp [d\lambda \cdot \frac D{d\lambda
}])v\right] _{(\tau _0+d\tau ,\lambda _0)}=  \nonumber \\
&=&\left[ (\exp [d\lambda \cdot \frac D{d\lambda }])(\exp [d\tau \cdot \frac
D{d\tau }])v\right] _{(\tau _0,\lambda _0)}\text{ \thinspace \thinspace
\thinspace \thinspace \thinspace .}  \label{1.47}
\end{eqnarray}

Now, we can compare the vectors $\widetilde{\widetilde{u}}$ and $\overline{%
\overline{u}}$ as well as $\widetilde{\widetilde{\xi }}$ and $\overline{%
\overline{\xi }}$ with respect to the vectors $u$ and $\xi $
correspondingly. For $d\tau =\delta \lambda =\varepsilon \ll 1$, we have 
\begin{equation}
\overline{\overline{u}}=\left[ (\exp [\varepsilon \cdot \frac D{d\lambda
}])(\exp [\varepsilon \cdot \frac D{d\tau }])u\right] _{(\tau _0,\lambda
_0)}\,\,\,\,\,\,\,\,\,\,\text{,}  \label{1.48a}
\end{equation}
\begin{equation}
\overline{\overline{\xi }}=\left[ (\exp [\varepsilon \cdot \frac D{d\lambda
}])(\exp [\varepsilon \cdot \frac D{d\tau }])\xi \right] _{(\tau _0,\lambda
_0)}\,\,\,\,\,\,\,\,\,\,\text{,}  \label{1.48b}
\end{equation}
\begin{equation}
\overline{\overline{v}}=\left[ (\exp [\varepsilon \cdot \frac D{d\lambda
}])(\exp [\varepsilon \cdot \frac D{d\tau }])v\right] _{(\tau _0,\lambda
_0)}\,\,\,\,\,\,\,\text{,}  \label{1.48c}
\end{equation}
\begin{equation}
\widetilde{\widetilde{u}}=\left[ (\exp [\varepsilon \cdot \frac D{d\tau
}])(\exp [\varepsilon \cdot \frac D{d\lambda }])u\right] _{(\tau _0,\lambda
_0)}\,\,\,\,\,\,\,\,\text{,}  \label{1.48d}
\end{equation}
\begin{equation}
\widetilde{\widetilde{\xi }}=\left[ (\exp [\varepsilon \cdot \frac D{d\tau
}])(\exp [\varepsilon \cdot \frac D{d\lambda }])\xi \right] _{(\tau
_0,\lambda _0)}\,\,\,\,\,\,\text{\thinspace \thinspace \thinspace ,}
\label{1.48e}
\end{equation}
\begin{equation}
\widetilde{\widetilde{v}}=\left[ (\exp [\varepsilon \cdot \frac D{d\tau
}])(\exp [\varepsilon \cdot \frac D{d\lambda }])v\right] _{(\tau _0,\lambda
_0)}\,\,\,\,\,\,\,\,\text{.}  \label{1.48f}
\end{equation}

Up to the second order, we obtain for the vectors $\overline{\overline{u}}$, 
$\overline{\overline{\xi }}$, and $\overline{\overline{v}}$ and for the
vectors $\widetilde{\widetilde{u}}$, $\widetilde{\widetilde{\xi }}$, and $%
\widetilde{\widetilde{v}}$%
\begin{eqnarray}
\overline{\overline{u}} &=&u_{(\tau _0,\lambda _0)}+\varepsilon \cdot (\frac{%
Du}{d\tau }+\frac{Du}{d\lambda })_{(\tau _0,\lambda _0)}+\varepsilon ^2\cdot
\left[ (\frac D{d\lambda }\circ \frac D{d\tau })u\right] _{(\tau _0,\lambda
_0)}+  \nonumber \\
&&+\frac 12\cdot \varepsilon ^2\cdot (\frac{D^2u}{d\lambda ^2}+\frac{D^2u}{%
d\tau ^2})_{(\tau _0,\lambda _0)}+O(\varepsilon ^3)\text{ ,}  \label{1.49}
\end{eqnarray}
\begin{eqnarray}
\widetilde{\widetilde{u}} &=&u_{(\tau _0,\lambda _0)}+\varepsilon \cdot (%
\frac{Du}{d\lambda }+\frac{Du}{d\tau })_{(\tau _0,\lambda _0)}+\varepsilon
^2\cdot \left[ (\frac D{d\tau }\circ \frac D{d\lambda })u\right] _{(\tau
_0,\lambda _0)}+  \nonumber \\
&&+\frac 12\cdot \varepsilon ^2\cdot (\frac{D^2u}{d\lambda ^2}+\frac{D^2u}{%
d\tau ^2})_{(\tau _0,\lambda _0)}+O(\varepsilon ^3)\text{ .}  \label{1.50}
\end{eqnarray}

The difference between the vectors $\overline{\overline{u}}$ and $\widetilde{%
\widetilde{u}}$ (transported along different paths from point $A$ to point $%
D $) can be found up to the second order of $\varepsilon $ in the form 
\begin{equation}
\overline{\overline{u}}-\widetilde{\widetilde{u}}=\varepsilon ^2\cdot \left[
(\frac D{d\lambda }\circ \frac D{d\tau }-\frac D{d\tau }\circ \frac
D{d\lambda })u\right] _{(\tau _0,\lambda _0)}+O(\varepsilon ^3)\,\,\,\,\,%
\text{,}  \label{1.51}
\end{equation}

\noindent or in the form 
\begin{equation}
\widetilde{\widetilde{u}}-\overline{\overline{u}}=\varepsilon ^2\cdot \left[
(\frac D{d\tau }\circ \frac D{d\lambda }-\frac D{d\lambda }\circ \frac
D{d\tau })u\right] _{(\tau _0,\lambda _0)}+O(\varepsilon ^3)\,\,\,\,\,\,%
\text{.}  \label{1.51a}
\end{equation}

The last expression could also be written in the form (having in mind that $%
D/d\tau =\nabla _u$ and $D/d\lambda =\nabla _\xi $) 
\[
\widetilde{\widetilde{u}}-\overline{\overline{u}}=\varepsilon ^2\cdot \left[
(\nabla _u\nabla _\xi -\nabla _\xi \nabla _u)u\right] _{(\tau _0,\lambda
_0)}+O(\varepsilon ^3)\,\,\,\,\,\,\,\,\,\text{,} 
\]
\begin{equation}
\overline{\overline{u}}-\widetilde{\widetilde{u}}=\varepsilon ^2\cdot \left[
(\nabla _\xi \nabla _u-\nabla _u\nabla _\xi )u\right] _{(\tau _0,\lambda
_0)}+O(\varepsilon ^3)\,\,\,\,\,\,\,\text{.}\,\,  \label{1.52}
\end{equation}

Since $\nabla _\xi \nabla _u-\nabla _u\nabla _\xi =R(\xi ,u)+\nabla
_{\pounds _\xi u}$, we obtain 
\begin{equation}
\overline{\overline{u}}-\widetilde{\widetilde{u}}=\varepsilon ^2\cdot
\{[R(\xi ,u)]u+\nabla _{\pounds _\xi u}u\}_{(\tau _0,\lambda
_0)}+O(\varepsilon ^3)\,\,\,\,\,\,\,\text{.}  \label{1.53}
\end{equation}

For a given vector field $v\in T(M)$ we obtain in an analogous relation 
\begin{equation}
\overline{\overline{v}}-\widetilde{\widetilde{v}}=\varepsilon ^2\cdot
\{[R(\xi ,u)]v+\nabla _{\pounds _\xi u}v\}_{(\tau _0,\lambda
_0)}+O(\varepsilon ^3)\,\,\,\,\,\,\,\text{.}  \label{1.54}
\end{equation}

Therefore, the curvature operator acting on a contravariant vector field $v$
determines the difference between the vectors, obtained by the transport of
the vector $v$ along two different paths constructing an infinitesimal
closed contour. If the curves $x^i(\tau ,\lambda _0)$ and $x^i(\tau
_0,\lambda )$ are co-ordinate lines then $\pounds _\xi u=0$ and therefore,
for an infinitesimal closed co-ordinate contour we have 
\begin{equation}
\overline{\overline{v}}-\widetilde{\widetilde{v}}=\varepsilon ^2\cdot
\{[R(\xi ,u)]v\}_{(\tau _0,\lambda _0)}+O(\varepsilon ^3)\,\,\,\,\,\,\,\,%
\text{,\thinspace \thinspace \thinspace \thinspace \thinspace \thinspace
\thinspace \thinspace \thinspace \thinspace }\pounds _\xi u=0\,\,\,\,\,\,%
\text{.\thinspace \thinspace \thinspace \thinspace }\,  \label{1.55}
\end{equation}

We have obtained (in a covariant manner) a well known result leading to the
geometrical interpretation of the curvature tensor in $[R(\xi ,u)]v$ as a
measure for the difference between the results of the different transports
(on different paths) of one and the same vector from one to an other point
of a closed contour.

\subsection{Covariant transport of a vector along a closed infinitesimal
contour from one and to the same point of the contour}

Let us now consider the transport of a contravariant vector $v$ from point $%
A $ along the closed infinitesimal contour $ACDBA$ to the same point $A$.

At the point $A$ we the vector $v$ could be denoted as $v_{(\tau _0,\lambda
_0)}$. At point $B$ the transported vector $v_{(\tau _0,\lambda _0+d\lambda
)}$ could be represented by means of the vector $v_{(\tau _0,\lambda _0)}$
as 
\begin{equation}
\stackunder{from\,\,\,p.\,\,A}{B}:\widetilde{v}=v_{(\tau _0,\lambda
_0+d\lambda )}=\left[ (\exp [d\lambda \cdot \frac D{d\lambda }])v\right]
_{(\tau _0,\lambda _0)}\text{ .}  \label{1.56}
\end{equation}

At the point $D$, the transported from point $B$ vector $v_{(\tau _0,\lambda
_0+d\lambda )}$ could be represented in the form 
\begin{eqnarray}
\stackunder{from\,\,p.\,B}{D} &:&\widetilde{\widetilde{v}}=v_{(\tau _0+d\tau
,\lambda _0+d\lambda )}=\left[ (\exp [d\tau \cdot \frac D{d\tau }])v\right]
_{(\tau _0,\lambda _0+d\lambda )}=  \nonumber \\
&=&\left[ (\exp [d\tau \cdot \frac D{d\tau }])(\exp [d\lambda \cdot \frac
D{d\lambda }])v\right] _{(\tau _0,\lambda _0)}\,\,\,\,\,\text{.}
\label{1.57}
\end{eqnarray}

At point $C$, the transported from point $D$ vector $v_{(\tau _0+d\tau
,\lambda _0+d\lambda )}$ can be written in the forms 
\begin{eqnarray}
\stackunder{from\,\,\,p.\,\,\,D}{C} &:&v_{ABDC}=v_{(\tau _0+d\tau ,\lambda
_0)}=\left[ (\exp [-d\lambda \cdot \frac D{d\lambda }])v\right] _{(\tau
_0+d\tau ,\lambda _0+d\lambda )}=  \nonumber \\
&=&\left[ (\exp [-d\lambda \cdot \frac D{d\lambda }])(\exp [d\tau \cdot
\frac D{d\tau }])v\right] _{(\tau _0,\lambda _0+d\lambda )}  \nonumber \\
&=&\left[ (\exp [-d\lambda \cdot \frac D{d\lambda }])(\exp [d\tau \cdot
\frac D{d\tau }])(\exp [d\lambda \cdot \frac D{d\lambda }])v\right] _{(\tau
_0,\lambda _0)}\text{\thinspace \thinspace \thinspace \thinspace \thinspace
\thinspace \thinspace .}  \label{1.58}
\end{eqnarray}

At point $A$, the transported from point $C$ vector $v_{(\tau _0+d\tau
,\lambda _0)}$ can be represented in the forms 
\begin{eqnarray}
\stackunder{from\,\,p.\,\,C}{A} &:&v_{ABDCA}=\left[ (\exp [-d\tau \cdot
\frac D{d\tau }])v\right] _{(\tau _0+d\tau ,\lambda _0)}=  \nonumber \\
&=&\left[ (\exp [-d\tau \cdot \frac D{d\tau }])(\exp [-d\lambda \cdot \frac
D{d\lambda }])(\exp [d\tau \cdot \frac D{d\tau }])(\exp [d\lambda \cdot
\frac D{d\lambda }])v\right] _{(\tau _0,\lambda _0)}\,\,\,\text{.}
\label{1.59}
\end{eqnarray}

If we use the explicit form of the exponent of the covariant differential
with $d\tau =d\lambda =\varepsilon $, we can find the expression for $%
v_{ABDCA}$ in the form 
\begin{eqnarray}
v_{ABDCA} &=&\{[1-\varepsilon \cdot \frac D{d\tau }+\frac 12\cdot
\varepsilon ^2\cdot \frac{D^2}{d\tau ^2}+\cdots ]\circ [1-\varepsilon \cdot
\frac D{d\lambda }+\frac 12\cdot \varepsilon ^2\cdot \frac{D^2}{d\lambda ^2}%
+\cdots ]\circ  \nonumber \\
&&\circ [1+\varepsilon \cdot \frac D{d\tau }+\frac 12\cdot \varepsilon
^2\cdot \frac{D^2}{d\tau ^2}+\cdots ]\circ [1+  \label{1.60} \\
&&+\varepsilon \cdot \frac D{d\lambda }+\frac 12\cdot \varepsilon ^2\cdot 
\frac{D^2}{d\lambda ^2}+\cdots ]v\}_{(\tau _0,\lambda _0)}\text{ .} 
\nonumber
\end{eqnarray}

Up to the second order of $\varepsilon $, we obtain the expressions 
\begin{eqnarray*}
&&\{[(1-\varepsilon \cdot \frac D{d\tau }+\frac 12\cdot \varepsilon ^2\cdot 
\frac{D^2}{d\tau ^2}-\varepsilon \cdot \frac D{d\lambda }+\varepsilon
^2\cdot \frac D{d\tau }\circ \frac D{d\lambda }+\frac 12\cdot \varepsilon
^2\cdot \frac{D^2}{d\lambda ^2}+O(\varepsilon ^3))\circ \\
&&\circ (1+\varepsilon \cdot \frac D{d\tau }+\frac 12\cdot \varepsilon
^2\cdot \frac{D^2}{d\tau ^2}+\varepsilon \cdot \frac D{d\lambda
}+\varepsilon ^2\cdot \frac D{d\tau }\circ \frac D{d\lambda }+\frac 12\cdot
\varepsilon ^2\cdot \frac{D^2}{d\lambda ^2}+O(\varepsilon ^3))]v\}_{(\tau
_0,\lambda _0)}= \\
&=&\{[1-\varepsilon \cdot \frac D{d\tau }+\frac 12\cdot \varepsilon ^2\cdot 
\frac{D^2}{d\tau ^2}-\varepsilon \cdot \frac D{d\lambda }+\varepsilon
^2\cdot \frac D{d\tau }\circ \frac D{d\lambda }+\frac 12\cdot \varepsilon
^2\cdot \frac{D^2}{d\lambda ^2}+ \\
&&+\varepsilon \cdot \frac D{d\tau }+\frac 12\cdot \varepsilon ^2\cdot \frac{%
D^2}{d\tau ^2}+\varepsilon \cdot \frac D{d\lambda }+\varepsilon ^2\cdot
\frac D{d\tau }\circ \frac D{d\lambda }+\frac 12\cdot \varepsilon ^2\cdot 
\frac{D^2}{d\lambda ^2}- \\
&&-\varepsilon ^2\cdot \frac{D^2}{d\tau ^2}-\varepsilon ^2\cdot \frac
D{d\lambda }\circ \frac D{d\tau }-\varepsilon ^2\cdot \frac D{d\tau }\circ
\frac D{d\lambda }-\varepsilon ^2\cdot \frac{D^2}{d\lambda ^2}+O(\varepsilon
^3)]v\}_{(\tau _0,\lambda _0)}
\end{eqnarray*}
\begin{equation}
v_{ABDCA}=\{[1+\varepsilon ^2\cdot \frac D{d\tau }\circ \frac D{d\lambda
}-\varepsilon ^2\cdot \frac D{d\lambda }\circ \frac D{d\tau }+O(\varepsilon
^3)]v\}_{(\tau _0,\lambda _0)}\,\,\text{,}  \label{1.61}
\end{equation}
\begin{equation}
v_{ABDCA}\approx v_{(\tau _0,\lambda _0)}+\varepsilon ^2\cdot \left\{ [\frac
D{d\tau }\circ \frac D{d\lambda }-\frac D{d\lambda }\circ \frac D{d\tau
}]v\right\} _{(\tau _0,\lambda _0)}\text{ \thinspace \thinspace \thinspace
\thinspace .}  \label{1.62}
\end{equation}

The difference between the vector $v_{(\tau _0,\lambda _0)}$ and the vector $%
v_{ABDCA}$, transported along the contour $ABDCA$ from the point $A$ to the
same point $A$, can be found up to the second order of $\varepsilon $ as 
\begin{equation}
v_{ABDCA}-v_{(\tau _0,\lambda _0)}\approx \varepsilon ^2\cdot \left\{ [\frac
D{d\tau }\circ \frac D{d\lambda }-\frac D{d\lambda }\circ \frac D{d\tau
}]v\right\} _{(\tau _0,\lambda _0)}\text{ \thinspace \thinspace \thinspace .}
\label{1.63}
\end{equation}

Since 
\begin{eqnarray}
\frac D{d\tau } &=&\nabla _u\,\,\,\,\,\,\,\text{,\thinspace \thinspace
\thinspace \thinspace \thinspace \thinspace \thinspace \thinspace \thinspace 
}u=\frac d{d\tau }\text{ \thinspace \thinspace \thinspace \thinspace
\thinspace ,\thinspace \thinspace \thinspace \thinspace \thinspace
\thinspace \thinspace \thinspace }\frac D{d\lambda }=\nabla _\xi \text{
\thinspace \thinspace \thinspace \thinspace ,\thinspace \thinspace
\thinspace \thinspace \thinspace \thinspace \thinspace }\xi =\frac
d{d\lambda }\text{ \thinspace \thinspace \thinspace ,}  \nonumber \\
\lbrack \frac D{d\tau }\circ \frac D{d\lambda }-\frac D{d\lambda }\circ
\frac D{d\tau }]v &=&[\nabla _u\nabla _\xi -\nabla _\xi \nabla _u]v=[R(u,\xi
)+\nabla _{\pounds _u\xi }]v=  \nonumber \\
&=&[R(u,\xi )]v+\nabla _{\pounds _u\xi }v=-[R(\xi ,u)]v-\nabla _{\pounds
_\xi u}v\text{ ,}  \label{1.64}
\end{eqnarray}

\noindent it follows that 
\begin{eqnarray}
v_{ABDCA} &\approx &v_{(\tau _0,\lambda _0)}-\varepsilon ^2\cdot \{[R(\xi
,u)]v-\nabla _{\pounds _\xi u}v\}_{(\tau _0,\lambda _0)}\text{ \thinspace
\thinspace ,}  \label{1.65a} \\
v_{(\tau _0,\lambda _0)}-v_{ABDCA} &\approx &\varepsilon ^2\cdot \{[R(\xi
,u)]v-\nabla _{\pounds _\xi u}v\}_{(\tau _0,\lambda _0)}\text{ .}
\label{1.65b}
\end{eqnarray}

Therefore, up to the second order of $\varepsilon $ we have 
\begin{eqnarray}
v_{(\tau _0,\lambda _0)}-v_{ABDCA} &=&\overline{\overline{v}}-\widetilde{%
\widetilde{v}}=v_{ACD}-v_{ABD}\text{ \thinspace \thinspace \thinspace ,}
\label{1.66} \\
v_{ACD} &=&\overline{\overline{v}}\,\,\,\,\,\,\text{,\thinspace \thinspace
\thinspace \thinspace \thinspace \thinspace \thinspace \thinspace }v_{ABD}=%
\widetilde{\widetilde{v}}\,\,\,\,\,\text{.}  \nonumber
\end{eqnarray}

The curvature tensor in $[R(\xi ,u)]v$ could also be interpreted as a
measure for the deviation of a vector, transported along a infinitesimal
closed contour, from the vector remaining at the same point from which the
transport began.

In an analogous way, we obtain for a covariant vector field $p\in T^{*}(M)$
the relation 
\begin{equation}
p_{(\tau _0,\lambda _0)}-p_{ABDCA}\approx \varepsilon ^2\cdot \{[R(\xi
,u)]p-\nabla _{\pounds _\xi u}p\}_{(\tau _0,\lambda _0)}\text{ .}
\label{1.67}
\end{equation}

\subsection{Action of the contraction operator $S$ on the pairs $%
(v,p)_{(\tau _0,\lambda _0)}$ and $(v_{ABDCA},p_{ABDCA})$}

From the relations 
\begin{eqnarray}
v_{(\tau _0,\lambda _0)} &\approx &v_{ABDCA}+\varepsilon ^2\cdot \{[R(\xi
,u)]v-\nabla _{\pounds _\xi u}v\}_{(\tau _0,\lambda _0)}\,\,\,\,\,\text{,}
\label{1.68a} \\
p_{(\tau _0,\lambda _0)} &\approx &p_{ABDCA}+\varepsilon ^2\cdot \{[R(\xi
,u)]p-\nabla _{\pounds _\xi u}p\}_{(\tau _0,\lambda _0)}\text{ \thinspace ,}
\label{1.68b}
\end{eqnarray}

\noindent it follows the result of the action of the contraction operator $S$
in the form 
\begin{eqnarray}
S(v,p)_{(\tau _0,\lambda _0)} &=&S\left( 
\begin{array}{c}
v_{ABDCA}+\varepsilon ^2\cdot \{[R(\xi ,u)]v-\nabla _{\pounds _\xi
u}v\}_{(\tau _0,\lambda _0)}, \\ 
p_{ABDCA}+\varepsilon ^2\cdot \{[R(\xi ,u)]p-\nabla _{\pounds _\xi
u}p\}_{(\tau _0,\lambda _0)}
\end{array}
\right) =  \nonumber \\
&=&S\left( v_{ABDCA},p_{ABDCA}\right) +  \nonumber \\
&&+\varepsilon ^2\cdot S\left( \{[R(\xi ,u)]v-\nabla _{\pounds _\xi
u}v\}_{(\tau _0,\lambda _0)},\,\,\,p_{ABDCA}\right) +  \nonumber \\
&&+\varepsilon ^2\cdot S\left( v_{ABDCA},\,\,\,\{[R(\xi ,u)]p-\nabla
_{\pounds _\xi u}p\}_{(\tau _0,\lambda _0)}\right) +  \nonumber \\
&&+\varepsilon ^4\cdot S\left( \{[R(\xi ,u)]v-\nabla _{\pounds _\xi
u}v\}_{(\tau _0,\lambda _0)},\,\,\,\,\,\{[R(\xi ,u)]p-\nabla _{\pounds _\xi
u}p\}_{(\tau _0,\lambda _0)}\right) \text{ \thinspace .}  \label{1.69}
\end{eqnarray}

For an infinitesimal closed contour, constructed from co-ordinate lines with
tangent vectors $u$ and $\xi $, the relation $\pounds _\xi u=-\pounds _u\xi
=0$ is valid and the above relations up to the second order of $\varepsilon $
have a simpler form 
\begin{eqnarray}
S(v,p)_{(\tau _0,\lambda _0)} &\approx &S\left( v_{ABDCA}+\varepsilon
^2\cdot \{[R(\xi ,u)]v\}_{(\tau _0,\lambda _0)},\,\,p_{ABDCA}+\varepsilon
^2\cdot \{[R(\xi ,u)]p\}_{(\tau _0,\lambda _0)}\right) =  \nonumber \\
&\approx &S\left( v_{ABDCA},p_{ABDCA}\right) +\varepsilon ^2\cdot S\left(
\{[R(\xi ,u)]v\}_{(\tau _0,\lambda _0)},\,\,\,p_{ABDCA}\right) +  \nonumber
\\
&&+\varepsilon ^2\cdot S\left( v_{ABDCA},\,\,\,\{[R(\xi ,u)]p\}_{(\tau
_0,\lambda _0)}\right) \,\,\,\text{,}  \label{1.70}
\end{eqnarray}
\[
S\left( v_{ABDCA},p_{ABDCA}\right) \approx S(v,p)_{(\tau _0,\lambda _0)}- 
\]
\begin{equation}
-\varepsilon ^2\cdot \left\{ S\left( \{[R(\xi ,u)]v\}_{(\tau _0,\lambda
_0)},\,\,\,p_{ABDCA}\right) +S\left( v_{ABDCA},\,\,\,\{[R(\xi ,u)]p\}_{(\tau
_0,\lambda _0)}\right) \right\} \,\,\,\text{.}  \label{1.71}
\end{equation}

On the other side, because of the relations 
\begin{eqnarray}
v_{ABDCA} &\approx &v_{(\tau _0,\lambda _0)}-\varepsilon ^2\cdot \{[R(\xi
,u)]v\}_{(\tau _0,\lambda _0)}\,\,\,\,\text{,}  \label{1.72a} \\
p_{ABDCA} &\approx &p_{(\tau _0,\lambda _0)}-\varepsilon ^2\cdot \{[R(\xi
,u)]p\}_{(\tau _0,\lambda _0)}\,\,\,\,\,\,\,\text{,}  \label{1.72b}
\end{eqnarray}

\noindent we obtain for $S\left( v_{ABDCA},p_{ABDCA}\right) $ up to the
second order of $\varepsilon $%
\begin{eqnarray}
S\left( v_{ABDCA},p_{ABDCA}\right) &\approx &S(v,p)_{(\tau _0,\lambda _0)}- 
\nonumber \\
&&-\varepsilon ^2\cdot \left\{ S\left( [R(\xi ,u)]v,\,\,\,p\right) +S\left(
v,\,\,\,[R(\xi ,u)]p\right) \right\} _{(\tau _0,\lambda _0)}\text{
\thinspace \thinspace \thinspace \thinspace .}  \label{1.73}
\end{eqnarray}

In a co-ordinate basis, the contravariant vector $[R(\xi ,u)]v$ and the
covariant vector $\,[R(\xi ,u)]p$ could be written in the forms \cite
{Manoff-0} 
\begin{eqnarray}
\lbrack R(\xi ,u)]v &=&R^l\,_{kij}\cdot \xi ^i\cdot u^j\cdot v^k\cdot
\partial _l\text{ \thinspace \thinspace \thinspace ,}  \label{1.74a} \\
\lbrack R(\xi ,u)]p &=&p_k\cdot P^k\,_{mij}\cdot \xi ^i\cdot u^j\cdot dx^m%
\text{ \thinspace \thinspace .}  \label{1.74b}
\end{eqnarray}

By the use of the last expressions for $[R(\xi ,u)]v$ and $[R(\xi ,u)]p$ in
the expressions for $S\left( v_{ABDCA},p_{ABDCA}\right) $, it follows that 
\begin{eqnarray}
S\left( v_{ABDCA},p_{ABDCA}\right) &\approx &S(v,p)_{(\tau _0,\lambda _0)}- 
\nonumber \\
&&-\varepsilon ^2\cdot [R^l\,_{kij}\cdot \xi ^i\cdot u^j\cdot v^k\cdot
S(\partial _l,p)+  \nonumber \\
&&+p_k\cdot P^k\,_{mij}\cdot \xi ^i\cdot u^j\cdot S(v,dx^m)]_{(\tau
_0,\lambda _0)}\text{ \thinspace .}  \label{1.75}
\end{eqnarray}

Since 
\begin{eqnarray}
S(\partial _l,p) &=&S(\partial _l,p_m\cdot dx^m)=p_m\cdot S(\partial
_l,dx^m)=p_m\cdot f^m\,_l\,\,\,\,\,\,\text{,}  \label{1.76a} \\
S(v,dx^m) &=&S(v^n\cdot \partial _n,dx^m)=v^n\cdot S(\partial
_n,dx^m)=v^n\cdot f^m\,_n\,\,\,\,\text{,}  \label{1.76b}
\end{eqnarray}

\noindent we can find the relations 
\begin{eqnarray}
S\left( v_{ABDCA},p_{ABDCA}\right) &\approx &S(v,p)_{(\tau _0,\lambda _0)}- 
\nonumber \\
&&-\varepsilon ^2\cdot [R^l\,_{kij}\cdot f^m\,_l\cdot \xi ^i\cdot u^j\cdot
v^k\cdot p_m+  \nonumber \\
&&+P^k\,_{mij}\cdot f^m\,_n\cdot \xi ^i\cdot u^j\cdot v^n\cdot p_k]_{(\tau
_0,\lambda _0)}\text{ \thinspace \thinspace ,}  \label{1.77}
\end{eqnarray}
\begin{eqnarray}
S\left( v_{ABDCA},p_{ABDCA}\right) &\approx &S(v,p)_{(\tau _0,\lambda _0)}- 
\nonumber \\
&&-\varepsilon ^2\cdot [R^{\overline{m}}\,_{kij}\cdot \xi ^i\cdot u^j\cdot
v^k\cdot p_m+  \nonumber \\
&&+P^k\,_{\overline{n}ij}\cdot \xi ^i\cdot u^j\cdot v^n\cdot p_k]_{(\tau
_0,\lambda _0)}\,\,\,\,\text{,}  \label{1.78}
\end{eqnarray}
\begin{eqnarray}
S\left( v_{ABDCA},p_{ABDCA}\right) &\approx &S(v,p)_{(\tau _0,\lambda _0)}- 
\nonumber \\
&&-\varepsilon ^2\cdot [(R^{\overline{k}}\,_{lij}+P^k\,_{\overline{l}%
ij})\cdot \xi ^i\cdot u^j\cdot v^l\cdot p_k]_{(\tau _0,\lambda _0)}\text{
\thinspace \thinspace \thinspace .}  \label{1.79}
\end{eqnarray}

On the other side, we have the integrability condition for the existence of
the contraction operator $S$ \cite{Manoff-0} 
\begin{equation}
R^{\overline{k}}\,_{lij}+P^k\,_{\overline{l}ij}=0\text{ \thinspace .}
\label{1.80}
\end{equation}

Because of the integrability condition, the last term of the right of the
expression for $S\left( v_{ABDCA},p_{ABDCA}\right) $ is equal to zero and it
follows that 
\begin{equation}
S\left( v_{ABDCA},p_{ABDCA}\right) \approx S(v,p)_{(\tau _0,\lambda
_0)}\,\,\,\,\,\text{.}  \label{1.81}
\end{equation}

This means that the result of the action of the contraction operator $S$ on
a contravariant and covariant vectors does not change after the transport of
both the vectors along an infinitesimal closed co-ordinate contour
(quadrangle).

In an analogous way, for the pairs $(\overline{\overline{v}},\overline{%
\overline{p}})$ and $(\widetilde{\widetilde{v}},\widetilde{\widetilde{p}})$,
transported on different paths from point $A$ to point $D$ of an
infinitesimal closed co-ordinate contour $ABDCA$, we obtain the relation 
\begin{equation}
S(\overline{\overline{v}},\overline{\overline{p}})_{(\tau _0+d\tau ,\lambda
_0+d\lambda )}=S(\widetilde{\widetilde{v}},\widetilde{\widetilde{p}})_{(\tau
_0+d\tau ,\lambda _0+d\lambda )}\text{ \thinspace \thinspace \thinspace
\thinspace \thinspace \thinspace ,\thinspace \thinspace \thinspace
\thinspace \thinspace \thinspace \thinspace \thinspace \thinspace \thinspace
\thinspace }d\tau =d\lambda =\varepsilon \ll 1\,\,\,\,\,.  \label{1.92}
\end{equation}

This means that the result of the action of the contraction operator $S$ on
a contravariant and a covariant tensors does not depend on the path on which
theses tensors are transported from one to an other point of a closed
infinitesimal co-ordinate contour.

\section{Geometrical interpretation of the torsion vector}

\subsection{Infinitesimal covariant transports and torsion vector}

Let a congruence of two parametric curves $x^i(\tau ,\lambda )$ be given.
Let the pair of vectors $(u,\xi )$ be transported from point $A$ with
co-ordinates $x^i(\tau _0,\lambda _0)$ to point $C$ with co-ordinates $%
x^i(\tau _0+d\tau ,\lambda _0)$ and on the other side, from point $A$ with
co-ordinates $x^i(\tau _0,\lambda _0)$ to point $B$ with co-ordinates $%
x^i(\tau _0,\lambda _0+d\lambda )$. The question arises under which
conditions the vectors $(u,\xi )$, transported from point $A$ to point $C$
[where they will be the vectors $(\overline{u},\overline{\xi })$] will be
equal to the vectors $(u,\xi )$, transported to point $B$ [where they will
be the vectors $(\widetilde{u},\widetilde{\xi })$].

The vectors $\overline{u}$ and $\overline{\xi }$ at point $C$ could be
represented by means of the vectors $u$ and $\xi $ at point $A$ if we use
the exponent of the covariant differential operator. Then, 
\begin{eqnarray}
\overline{u} &:&=u_{(\tau _0+d\tau ,\lambda _0)}=\left( \exp [d\tau \cdot
\frac D{d\tau }]\right) u_{(\tau _0,\lambda _0)}\text{ ,}  \label{2.1a} \\
\overline{\xi } &:&=\xi _{(\tau _0+d\tau ,\lambda _0)}=\left( \exp [d\tau
\cdot \frac D{d\tau }]\right) \xi _{(\tau _0,\lambda _0)\text{ }}\,\text{.}
\label{2.1b}
\end{eqnarray}

At the same time, the vectors $\widetilde{u}$ and $\widetilde{\xi }$ at
point $C$ could also be represented by means of the vectors $u$ and $\xi $
at the point $A$ in the form: 
\begin{eqnarray}
\widetilde{u} &:&=u_{(\tau _0,\lambda _0+d\lambda )}=\left( \exp [d\lambda
\cdot \frac D{d\lambda }]\right) u_{(\tau _0,\lambda _0)}\text{ ,}
\label{2.2a} \\
\widetilde{\xi } &:&=\xi _{(\tau _0,\lambda _0+d\lambda )}=\left( \exp
[d\lambda \cdot \frac D{d\lambda }]\right) \xi _{(\tau _0,\lambda _0)\text{ }%
}\,\text{.}  \label{2.2b}
\end{eqnarray}

If the conditions 
\[
\overline{u}=\widetilde{u}\,\,\,\,\,\,\text{,\thinspace \thinspace
\thinspace \thinspace }\overline{\xi }=\widetilde{\xi }\text{\thinspace
\thinspace \thinspace \thinspace \thinspace \thinspace } 
\]

\noindent %
should be fulfilled for $d\tau =d\lambda =\varepsilon \ll 1$, then the
relations 
\begin{eqnarray}
\left( \exp [\varepsilon \cdot \frac D{d\tau }]\right) u_{(\tau _0,\lambda
_0)} &=&\left( \exp [\varepsilon \cdot \frac D{d\lambda }]\right) u_{(\tau
_0,\lambda _0)}\text{ \thinspace \thinspace \thinspace \thinspace \thinspace
\thinspace \thinspace \thinspace \thinspace \thinspace ,}  \label{2.3a} \\
\left( \exp [\varepsilon \cdot \frac D{d\tau }]\right) \xi _{(\tau
_0,\lambda _0)\text{ }} &=&\left( \exp [\varepsilon \cdot \frac D{d\lambda
}]\right) \xi _{(\tau _0,\lambda _0)\text{ }}\,\,\text{\thinspace \thinspace
\thinspace \thinspace \thinspace \thinspace \thinspace \thinspace ,}
\label{2.3b}
\end{eqnarray}

\noindent should be valid. The last conditions could be written in their
explicit forms 
\begin{eqnarray}
&&\left[ \left( 1+\varepsilon \cdot \frac D{d\tau }+\frac 1{2!}\cdot
\varepsilon ^2\cdot \frac{D^2}{d\tau ^2}+\cdots \right) u\right] _{(\tau
_0,\lambda _0)}\text{ }  \nonumber \\
&=&\left[ \left( 1+\varepsilon \cdot \frac D{d\lambda }+\frac 1{2!}\cdot
\varepsilon ^2\cdot \frac{D^2}{d\lambda ^2}+\cdots \right) u\right] _{(\tau
_0,\lambda _0)}\text{ \thinspace \thinspace \thinspace \thinspace ,}
\label{2.4a}
\end{eqnarray}
\begin{eqnarray}
&&\left[ \left( 1+\varepsilon \cdot \frac D{d\tau }+\frac 1{2!}\cdot
\varepsilon ^2\cdot \frac{D^2}{d\tau ^2}+\cdots \right) \xi \right] _{(\tau
_0,\lambda _0)}\text{ }  \nonumber \\
&=&\text{\thinspace \thinspace \thinspace }\left[ \left( 1+\varepsilon \cdot
\frac D{d\lambda }+\frac 1{2!}\cdot \varepsilon ^2\cdot \frac{D^2}{d\lambda
^2}+\cdots \right) \xi \right] _{(\tau _0,\lambda _0)}\text{ \thinspace
\thinspace \thinspace \thinspace .}  \label{2.4b}
\end{eqnarray}

Up to the first order of $\varepsilon $, we obtain the relations 
\begin{eqnarray}
\left( \varepsilon \cdot \frac{Du}{d\tau }\right) _{(\tau _0,\lambda _0)}
&=&\left( \varepsilon \cdot \frac{Du}{d\lambda }\right) _{(\tau _0,\lambda
_0)}\text{ \thinspace \thinspace \thinspace \thinspace ,\thinspace
\thinspace \thinspace \thinspace \thinspace \thinspace \thinspace \thinspace
\thinspace \thinspace }\varepsilon \neq 0\text{ ,}  \label{2.5a} \\
\left( \varepsilon \cdot \frac{D\xi }{d\tau }\right) _{(\tau _0,\lambda _0)}
&=&\left( \varepsilon \cdot \frac{D\xi }{d\lambda }\right) _{(\tau
_0,\lambda _0)}\text{ \thinspace \thinspace \thinspace \thinspace
,\thinspace \thinspace \thinspace \thinspace \thinspace \thinspace
\thinspace \thinspace \thinspace \thinspace }\varepsilon \neq 0\text{ ,}
\label{2.5b}
\end{eqnarray}

\noindent %
leading to the conditions [at the point $x^i(\tau _0,\lambda _0)$] 
\begin{eqnarray}
\nabla _uu &=&\nabla _\xi u\text{ \thinspace \thinspace \thinspace
\thinspace \thinspace \thinspace \thinspace \thinspace \thinspace \thinspace
,\thinspace \thinspace \thinspace \thinspace \thinspace \thinspace
\thinspace \thinspace \thinspace \thinspace \thinspace \thinspace \thinspace
\thinspace }u=\frac d{d\tau }=u^i\cdot \partial _i\text{ \thinspace
\thinspace ,\thinspace \thinspace \thinspace \thinspace \thinspace
\thinspace \thinspace \thinspace \thinspace \thinspace }u^i=\frac{dx^i}{%
d\tau }\text{ \thinspace \thinspace \thinspace ,\thinspace \thinspace
\thinspace \thinspace \thinspace }u\in T(M)\text{ \thinspace \thinspace ,}
\label{2.6a} \\
\nabla _u\xi &=&\nabla _\xi \xi \text{ \thinspace \thinspace \thinspace
\thinspace \thinspace \thinspace \thinspace \thinspace \thinspace \thinspace
,\thinspace \thinspace \thinspace \thinspace \thinspace \thinspace
\thinspace \thinspace \thinspace \thinspace \thinspace \thinspace \thinspace
\thinspace }\xi =\frac d{d\lambda }=\xi ^i\cdot \partial _i\text{ \thinspace
\thinspace \thinspace ,\thinspace \thinspace \thinspace \thinspace
\thinspace \thinspace \thinspace \thinspace }\xi ^i=\frac{dx^i}{d\lambda }%
\text{ \thinspace \thinspace \thinspace \thinspace ,\thinspace \thinspace
\thinspace \thinspace }\xi \in T(M)\text{ .}  \label{2.6b}
\end{eqnarray}

On the other side, the vectors $u$ and $\xi $ fulfill the relation 
\[
\pounds _\xi u=\nabla _\xi u-\nabla _u\xi -T(\xi ,u)\text{ \thinspace
\thinspace \thinspace \thinspace .} 
\]

Substituting $\nabla _\xi u$ and $\nabla _u\xi $ from the above conditions,
we obtain 
\begin{equation}
\pounds _\xi u=\nabla _uu-\nabla _\xi \xi -T(\xi ,u)\text{ }\,\,\,\,\,\text{%
.\thinspace }  \label{2.7}
\end{equation}

If the vectors $u$ and $\xi $ are tangent vectors to the co-ordinate lines $%
\tau $ and $\lambda $ correspondingly, then $\pounds _\xi u=0$ and the
relation 
\begin{equation}
\nabla _uu-\nabla _\xi \xi =T(\xi ,u)\text{ \thinspace \thinspace }%
\,\,\,\,\,\,\,\,\,\,\,\text{,\thinspace \thinspace \thinspace \thinspace
\thinspace \thinspace \thinspace \thinspace \thinspace \thinspace \thinspace
\thinspace \thinspace \thinspace }T(\xi ,u)=T_{ij}\,^k\cdot \xi ^i\cdot
u^j\cdot \partial _k\in T(M)\text{ ,}  \label{2.8}
\end{equation}

\noindent %
follows. This means that under the assumption of the equivalence of the
pairs $(\overline{u},\overline{\xi })$ and $(\widetilde{u},\widetilde{\xi })$
[obtained as a result of the transport of the pair $(u,\xi )$ from point $A$
to points $C$ and $B$ respectively] the torsion vector $T(\xi ,u)$ and the
corresponding torsion tensor $T=T_{ij}\,^k\cdot dx^i\wedge dx^j\otimes
\partial _k$ can be interpreted as a measure for the deviation of the
transport of a vector $u$ along itself from the transport of the vector $\xi 
$ along itself. It should be stressed that all vectors are compared to each
other at the beginning point $A$ of the transport, i.e. they are expressed
by means of the corresponding vectors, found at point $A$ by the use of the
exponent mapping (exponent of the covariant differential operator) acting on
the vectors $u$ and $\xi $ at point $A$.

\subsection{Infinitesimal covariant transport and Lie derivative}

Let us now consider the conditions for transports under which the Lie
derivative remains unchanged.

Let the Lie derivative $\pounds _\xi u$ be given at a point $A$ with
co-ordinates $x^i(\tau _0,\lambda _0)$%
\begin{equation}
\left[ \pounds _\xi u\right] _{(\tau _0,\lambda _0)}=[\xi ,u]_{(\tau
_0,\lambda _0)}\text{ \thinspace \thinspace \thinspace \thinspace \thinspace
\thinspace \thinspace \thinspace .}  \label{2.9}
\end{equation}

At point $B$, the transported vectors $\xi $ and $u$ will have the forms
respectively: 
\begin{eqnarray}
\widetilde{\xi } &:&=\xi _{(\tau _0,\lambda _0+d\lambda )}=\left( \exp
[d\lambda \cdot \frac D{d\lambda }]\right) \xi _{(\tau _0,\lambda _0)\text{ }%
}\text{ \thinspace \thinspace \thinspace ,}  \label{2.10a} \\
\widetilde{u} &:&=u_{(\tau _0,\lambda _0+d\lambda )}=\left( \exp [d\lambda
\cdot \frac D{d\lambda }]\right) u_{(\tau _0,\lambda _0)}\text{ \thinspace
\thinspace \thinspace \thinspace \thinspace \thinspace .}  \label{2.10b}
\end{eqnarray}

The Lie derivative $\pounds _{\widetilde{\xi }}\widetilde{u}$ at the point $%
B $ could be expressed as 
\begin{equation}
\lbrack \widetilde{\xi },\widetilde{u}]=\left[ \left( \exp [d\lambda \cdot
\frac D{d\lambda }]\right) \xi _{(\tau _0,\lambda _0)\text{ }},\,\,\left(
\exp [d\lambda \cdot \frac D{d\lambda }]\right) u_{(\tau _0,\lambda
_0)}\right] \text{ \thinspace \thinspace \thinspace \thinspace .}
\label{2.11}
\end{equation}

If we use the explicit form of the exponent of the covariant differential
operator for the case $d\tau =d\lambda =\varepsilon $ we can write $[%
\widetilde{\xi },\widetilde{u}]$ in the form 
\begin{eqnarray}
\lbrack \widetilde{\xi },\widetilde{u}] &=&\left[ 
\begin{array}{c}
\left[ \left( 1+\varepsilon \cdot \frac D{d\lambda }+\frac 1{2!}\cdot
\varepsilon ^2\cdot \frac{D^2}{d\lambda ^2}+\cdots \right) \xi \right]
_{(\tau _0,\lambda _0)}\,\,\,\, \\ 
,\,\,\left[ \left( 1+\varepsilon \cdot \frac D{d\lambda }+\frac 1{2!}\cdot
\varepsilon ^2\cdot \frac{D^2}{d\lambda ^2}+\cdots \right) u\right] _{(\tau
_0,\lambda _0)}
\end{array}
\right] \,=  \nonumber \\
&=&\left[ \left( 1+\varepsilon \cdot \frac D{d\lambda }+\frac 1{2!}\cdot
\varepsilon ^2\cdot \frac{D^2}{d\lambda ^2}+\cdots \right) \xi \right]
_{(\tau _0,\lambda _0)}  \nonumber \\
&&\circ \left[ \left( 1+\varepsilon \cdot \frac D{d\lambda }+\frac
1{2!}\cdot \varepsilon ^2\cdot \frac{D^2}{d\lambda ^2}+\cdots \right)
u\right] _{(\tau _0,\lambda _0)}-  \nonumber \\
&&-\left[ \left( 1+\varepsilon \cdot \frac D{d\lambda }+\frac 1{2!}\cdot
\varepsilon ^2\cdot \frac{D^2}{d\lambda ^2}+\cdots \right) u\right] _{(\tau
_0,\lambda _0)}  \nonumber \\
&&\circ \left[ \left( 1+\varepsilon \cdot \frac D{d\lambda }+\frac
1{2!}\cdot \varepsilon ^2\cdot \frac{D^2}{d\lambda ^2}+\cdots \right) \xi
\right] _{(\tau _0,\lambda _0)}\text{ \thinspace \thinspace \thinspace
\thinspace .}  \label{2.12}
\end{eqnarray}

Up to the first order of $\varepsilon $, we obtain 
\begin{eqnarray}
\lbrack \widetilde{\xi },\widetilde{u}] &\approx &[\xi ,u]_{(\tau _0,\lambda
_0)}+\varepsilon \cdot [\frac{D\xi }{d\lambda },u]_{(\tau _0,\lambda
_0)}+\varepsilon \cdot [\xi ,\frac{Du}{d\lambda }]_{(\tau _0,\lambda _0)}= 
\nonumber \\
&=&[\xi ,u]_{(\tau _0,\lambda _0)}+\varepsilon \cdot \left\{ [\frac{D\xi }{%
d\lambda },u]+[\xi ,\frac{Du}{d\lambda }]\right\} _{(\tau _0,\lambda _0)}%
\text{ \thinspace \thinspace \thinspace \thinspace \thinspace \thinspace
\thinspace \thinspace \thinspace \thinspace \thinspace .}  \label{2.13}
\end{eqnarray}

Since $[\xi ,u]=\pounds _\xi u$ is a vector, we have 
\begin{eqnarray}
\left( \nabla _\xi [\xi ,u]\right) _{(\tau _0,\lambda _0)} &=&\left( \frac
D{d\lambda }[\xi ,u]\right) _{(\tau _0,\lambda _0)}=\stackunder{\varepsilon
\rightarrow 0}{\lim }\frac{[\xi ,u]_{(\tau _0,\lambda _0+d\lambda )}-[\xi
,u]_{(\tau _0,\lambda _0)}}\varepsilon =  \nonumber \\
&=&\stackunder{\varepsilon \rightarrow 0}{\lim }\frac{[\widetilde{\xi },%
\widetilde{u}]-[\xi ,u]_{(\tau _0,\lambda _0)}}\varepsilon =\left\{ [\frac{%
D\xi }{d\lambda },u]+[\xi ,\frac{Du}{d\lambda }]\right\} _{(\tau _0,\lambda
_0)}\text{ \thinspace \thinspace \thinspace \thinspace \thinspace .}
\label{2.14}
\end{eqnarray}

Therefore, for every point of the curve $x^i(\tau ,\lambda )$ the relation 
\begin{eqnarray}
\frac D{d\lambda }[\xi ,u] &=&[\frac{D\xi }{d\lambda },u]+[\xi ,\frac{Du}{%
d\lambda }]\text{ }  \nonumber \\
\nabla _\xi [\xi ,u] &=&[\nabla _\xi \xi ,u]+[\xi ,\nabla _\xi u]\text{
\thinspace \thinspace \thinspace \thinspace \thinspace , \thinspace
\thinspace \thinspace \thinspace \thinspace \thinspace \thinspace \thinspace
\thinspace \thinspace \thinspace \thinspace }\xi =\frac d{d\lambda }\text{
\thinspace \thinspace \thinspace \thinspace \thinspace \thinspace ,}
\label{2.15}
\end{eqnarray}

\noindent %
is valid. Now we can write the expression 
\begin{equation}
\left( \lbrack \xi ,u]\right) _{\left( \tau _0,\lambda _0+\varepsilon
\right) }=\left( [\xi ,u]\right) _{(\tau _0,\lambda _0)}+\varepsilon \cdot
\left( \frac D{d\lambda }[\xi ,u]\right) _{(\tau _0,\lambda _0)}\text{
\thinspace \thinspace \thinspace \thinspace \thinspace \thinspace .}
\label{2.16}
\end{equation}

For every given pair of vector fields $(v,w)$, $v,w\in T(M)$ analogous
relations are fulfilled 
\begin{equation}
\left( \lbrack v,w]\right) _{\left( \tau _0,\lambda _0+\varepsilon \right)
}=\left( [v,w]\right) _{(\tau _0,\lambda _0)}+\varepsilon \cdot \left( \frac
D{d\lambda }[v,w]\right) _{(\tau _0,\lambda _0)}\text{ \thinspace \thinspace
\thinspace .}  \label{2.17}
\end{equation}

On the basis of the last relation the following proposition can be proved:

\begin{proposition}
The necessary and sufficient condition for not changing of the Lie
derivative of two vector fields $v$ and $w$ along a curve $x^i(\tau _0=$
const.$,\lambda )$ is the condition 
\begin{equation}
\frac D{d\lambda }[v,w]=0\text{ }\,\,\,\,\,\,\,\,\,\text{or \thinspace
\thinspace \thinspace \thinspace \thinspace }\nabla _\xi \pounds _vw=0\text{
,\thinspace \thinspace \thinspace \thinspace \thinspace \thinspace
\thinspace \thinspace \thinspace \thinspace }\xi =\frac d{d\lambda }\text{%
\thinspace \thinspace \thinspace \thinspace \thinspace .}  \label{2.18}
\end{equation}
\end{proposition}

Therefore, the Lie derivative $\pounds _vw=[v,w]$ of a contravariant vector
field $w\in T(M)$ along an other contravariant vector $v\in T(M)$ could not
change by an infinitesimal covariant transport from a point $A$ with
co-ordinates $x^i(\tau _0,\lambda _0)$ of a curve $x^i(\tau =\tau _0,\lambda
)$ to a point $B$ with co-ordinates $x^i(\tau =\tau _0,\lambda
_0+\varepsilon )$ of the same curve. This means that the necessary and
sufficient condition for not changing of the Lie derivative under an
infinitesimal transport from one to an other point of a curve is the
vanishing of the covariant derivative of the Lie derivative along the vector
field tangential to every point of the curve.

{\it Special case: }$\pounds _vw=0$, \thinspace $v$, $w\in T(M)$, $%
v:=d/d\rho $, $w:=d/d\sigma $. In this special case the vector field $v$ and 
$w$ are tangent vector fields to the corresponding co-ordinate lines $%
x^i(\rho ,\sigma =$ const.$)$ and $x^i(\rho ,\sigma =$ const.$)$. The
congruence of these co-ordinate lines should obey the relations 
\begin{equation}
\nabla _v\pounds _vw=0\text{ \thinspace \thinspace \thinspace \thinspace
\thinspace ,\thinspace \thinspace \thinspace \thinspace \thinspace
\thinspace \thinspace }\nabla _w\pounds _vw=0\text{ \thinspace \thinspace
\thinspace \thinspace ,}  \label{2.19}
\end{equation}

\noindent %
i.e. the contravariant vector field $\pounds _vw=[v,w]$ should be
transported parallel to the corresponding co-ordinate line.

If $\pounds _vw=[v,w]=0$ then $\nabla _vw-\nabla _wv=T(v,w)$. The torsion
vector $T(v,w)$ determines the difference between $\nabla _vw$ and $\nabla
_wv$. If one of the vector fields $w$ or $v$ is transported parallel to the
other, i.e. if $\nabla _vw=0$ or $\nabla _wv=0$ correspondingly, then the
torsion vector $T(v,w)\neq 0$ is an obstacle for the other vector field ($v$
or $w$ correspondingly) to be also transported parallel to $w$ or $v$
correspondingly.. Therefore, the torsion vector (and the torsion tensor)
could be interpreted as a measure for a deviation of parallel transport of a
vector field, when the other vector field is transported parallel to it and
at the same time both the vector fields are tangent vector fields to a
congruence of two parametric lines which could be used as co-ordinate lines
in a differentiable manifold $M$ . In other words, if two co-ordinate lines
in a space with torsion have tangent vectors, where the first of which could
be transported parallel to the second vector, then the second tangent vector
could not be transported parallel to the first one if the torsion vector is
different from zero. This fact could be used for description of physical
situations, where only one of the vector fields could be transported
parallel along a co-ordinate line. This could have influence on the
kinematic characteristics of a flow.

\section{Kinematic characteristics of a flow}

Let us now consider a parametric representation of an $n$-parametric ($n$%
-dimensional) congruence (family) of curves with respect to an arbitrary
co-ordinate system \cite{Ehlers} in an $n$-dimensional differentiable
manifold $M$ 
\begin{equation}
x^i=x^i(\tau ,\lambda ^a)\text{ , \ \ \ \ \ \ \ \ }i=1,2,....,n\text{,
\thinspace \ \ \ \ \ \ }a=1,2,...,n-1\text{, \ \ }(\dim M=n)\text{.\ }
\label{4.1}
\end{equation}

The parameters $\lambda ^a$ designate the matter elements (material points), 
$\tau $ is the parameter along the curves interpreted for $n=4$ as the
proper time along the world lines $x^i=x^i(\tau ,\lambda ^a=\,$const$.)$.
The curves with parameter $\tau $ could also be interpreted as the line's
having at every of its points tangential vector identified with the velocity
of a material point at this line's point. The set of all such lines is call
a flow. A single line of a flow is called line's flow. The tangent vector to
the curves $x^i=x^i(\tau ,\lambda ^a=\lambda _0^a=\,$const.$)$%
\begin{equation}
u:=\frac \partial {\partial \tau }=\frac d{d\tau }=\frac{\partial x^i}{%
\partial \tau }\cdot \partial _i=\frac{dx^i}{d\tau }\cdot \partial
_i=u^i\cdot \partial _i\text{ , \ \ \ \ \ \ \ \ }u^i=\frac{\partial x^i}{%
\partial \tau }=\frac{dx^i}{d\tau }\text{ \ ,}  \label{4.2}
\end{equation}

\noindent %
is interpreted as the velocity of the material points in the media. The
transformation $x^i:(\tau ,\lambda ^a)\rightarrow x^i(\tau ,\lambda ^a)$
corresponds to the transformation in classical mechanics from Lagrangian to
Eulerian co-ordinates \cite{Ehlers}.

Let $\xi _{(a)}$ be vectors, tangential to the curves $x^i(\tau =\,$const.$%
,\lambda ^a)$. The curves $x^i(\tau =\tau _0=\,$const.$,\lambda ^a)$
describe the positions of material points for a given parameter (proper
time) $\tau =\tau _0=$ const., i.e. 
\begin{eqnarray}
\xi _{(a)} &:&=\frac \partial {\partial \lambda ^a}=\frac d{d\lambda ^a}=%
\frac{\partial x^i}{\partial \lambda ^a}\cdot \partial _i=\xi _{(a)}^i\cdot
\partial _i\text{ \thinspace \thinspace \thinspace \thinspace \thinspace
\thinspace ,\thinspace \thinspace \thinspace \thinspace \thinspace
\thinspace \thinspace \thinspace }\xi _{(a)}^i:=\frac{\partial x^i}{\partial
\lambda ^a}=\frac{dx^i}{d\lambda ^a}\text{ \thinspace \thinspace \thinspace
\thinspace \thinspace ,\thinspace \thinspace \thinspace }  \label{4.3} \\
\text{\thinspace \thinspace \thinspace \thinspace \thinspace \thinspace }%
\frac{d\lambda ^a}{d\tau } &=&0\text{ ,\thinspace \thinspace \thinspace
\thinspace \thinspace }\frac{d\tau }{d\lambda ^a}=0\text{ \thinspace
\thinspace \thinspace .}  \nonumber
\end{eqnarray}

If we consider a neighborhood of a point $P$ of the flow with co-ordinates $%
x_P^i=x^i(\tau _0,\lambda _0^a)$ then we can find the co-ordinates of two
points from a neighborhood of the point $P$ denoted as point $P_1$ with
co-ordinates $x_{P_1}^i=x^i(\tau =\tau _0+d\tau ,\lambda _0^a)$ and point $%
P_2$ with co-ordinates $x_{P_2}^i=x^i(\tau =\tau _0,\lambda ^a=\lambda
_0^a+d\lambda )$, where $\tau _0=$ const.,\thinspace \thinspace $\lambda _0=$
const.: 
\begin{eqnarray}
x_{(\tau _0+d\tau ,\lambda _0^a)}^i &=&x_{(\tau _0,\lambda _0^a)}^i+d\tau
\cdot \left( \frac{\partial x^i}{\partial \tau }\right) _{(\tau _0,\lambda
_0^a)}+O(d\tau ^2)\approx  \nonumber \\
&\approx &x_{(\tau _0,\lambda _0^a)}^i+d\tau \cdot u_{(\tau _0,\lambda
_0^a)}^i=x_{(\tau _0,\lambda _0^a)}^i+\overline{u}_{(\tau _0,\lambda _0^a)}^i%
\text{ \thinspace \thinspace \thinspace \thinspace \thinspace \thinspace
,\thinspace }  \nonumber \\
\text{\thinspace \thinspace \thinspace \thinspace }\overline{u}_{(\tau
_0,\lambda _0^a)}^i &:&=d\tau \cdot u_{(\tau _0,\lambda _0^a)}^i\text{%
\thinspace \thinspace \thinspace \thinspace \thinspace ,}  \nonumber \\
x_{(\tau _0,\lambda _0^a+d\lambda ^a)}^i &=&x_{(\tau _0,\lambda
_0^a)}^i+d\lambda ^a\cdot \left( \frac{\partial x^i}{d\lambda ^a}\right)
_{(\tau _0,\lambda _0^a)}+O\left( (d\lambda ^a)^2\right) \approx  \label{4.4}
\\
&\approx &x_{(\tau _0,\lambda _0^a)}^i+d\lambda ^a\cdot \xi _{(a)(\tau
_0,\lambda _0)}^i=x_{(\tau _0,\lambda _0^a)}^i+\overline{\xi }_{(a)(\tau
_0,\lambda _0)}^i\text{ \thinspace \thinspace \thinspace \thinspace
,\thinspace \thinspace \thinspace }  \nonumber \\
\text{\thinspace \thinspace \thinspace \thinspace \thinspace }\overline{\xi }%
_{(a)(\tau _0,\lambda _0)}^i &:&=d\lambda ^a\cdot \xi _{(a)(\tau _0,\lambda
_0)}^i\text{ ,}  \nonumber \\
&&\text{(there is no summation over }a\text{).}  \nonumber
\end{eqnarray}

If the motion of a material point is described along the curve $x^i(\tau
,\lambda ^a=\lambda _0^a)$ from the point $P$ with co-ordinates $x^i(\tau
_0,\lambda _0^a)$ to the point $P_1$ with the co-ordinates $x^i(\tau
_0+d\tau ,\lambda _0^a)$ then the velocity of this material point along the
curve with parameter $\tau $ will be 
\begin{equation}
u^i=\stackunder{d\tau \rightarrow 0}{\lim }\frac{x^i(\tau _0+d\tau ,\lambda
_0^a)-x^i(\tau _0,\lambda _0^a)}{d\tau }\text{ .}  \label{4.5}
\end{equation}

Correspondingly, the velocity of a material point moving from the point $P$
[with co-ordinates $x^i(\tau _0,\lambda _0^a)$] to the point $P_2$ [with the
co-ordinates $x^i(\tau _0,\lambda _0^a+d\lambda ^a)$] along the curve with
parameter $\lambda $ will be 
\begin{equation}
\xi _{(a)}^i=\stackunder{d\lambda \rightarrow 0}{\lim }\frac{x^i(\tau
_0,\lambda _0^a+d\lambda ^a)-x^i(\tau _0,\lambda _0^a)}{d\lambda }\text{
\thinspace \thinspace \thinspace .}  \label{4.6}
\end{equation}

Therefore, the vectors $u$ and $\xi _{(a)}$ represent the velocity of
material points moving on a curve with parameter $\tau $ or on a curve with
parameter $\lambda ^a$ correspondingly. The notion of velocity along a curve
is not jet physically interpreted. If the parameter $\tau $ is interpreted
as the proper time of a material point moving on the curve $x^i(\tau
,\lambda _0^a=$ const.$)$ then the vector $u=u^i\cdot \partial _i$ is its
local velocity and the vector $\overline{\xi }_{(a)(\tau ,\lambda
_0^a+d\lambda ^a)}=d\lambda ^a\cdot \xi _{(a)(\tau ,\lambda _0^a+d\lambda
^a)}^i$ (there is no summation over $a$) is a distance vector of a material
point from the point $\overline{P}$ with co-ordinates $x^i(\tau ,\lambda
_0^a)$.

If $\tau $ and $\lambda ^a$ are chosen to be co-ordinates of the material
points of a flow [proper frame of reference (moving with the points of the
flow)] then the tangent vectors $u$ and $\xi _{(a)}$ should obey the
conditions \cite{Schutz}: 
\begin{equation}
\pounds _u\xi _{(a)}=0\text{ ,\thinspace \thinspace \thinspace \thinspace
\thinspace \thinspace \thinspace \thinspace \thinspace \thinspace \thinspace
\thinspace \thinspace \thinspace \thinspace \thinspace \thinspace \thinspace
\thinspace \thinspace }a=1,...,n-1\text{ \thinspace \thinspace \thinspace .}
\label{4.7}
\end{equation}

At the same time, the vectors $\xi _{(a)}$, $a=1,...,n-1$, should be
orthogonal to the vector $u$ (in order to be a linear independent vector set
in the manifold $M$ with $\dim M=n$), i.e. $\xi _{(a)}$ should obey the
relations 
\begin{equation}
g(u,\xi _{(a)})=0\text{ \thinspace \thinspace \thinspace \thinspace .}
\label{4.8}
\end{equation}

The last conditions mean that $\xi _{(a)}$ could be expressed in the form 
\begin{equation}
\xi _{(a)}=\xi _{(a)\perp }:=\overline{g}[h_u(\xi _{(a)})]\text{ ,}
\label{4.9}
\end{equation}

\noindent because of the decomposition of the vectors $\xi _{(a)}$ in a part
collinear to the vector $u$ and a part orthogonal to $u$, i.e.

\begin{equation}
\xi _{(a)}=\frac{l_a}e\cdot u+\xi _{(a)\perp }\text{ ,}  \label{4.10}
\end{equation}

\noindent where 
\begin{eqnarray}
l_a &:&=g(u,\xi _{(a)})\text{ \thinspace \thinspace \thinspace ,\thinspace
\thinspace \thinspace \thinspace \thinspace \thinspace \thinspace \thinspace
\thinspace \thinspace \thinspace \thinspace \thinspace }e:=g(u,u):\neq 0%
\text{ \thinspace \thinspace \thinspace \thinspace \thinspace \thinspace
\thinspace \thinspace ,\thinspace \thinspace \thinspace \thinspace
\thinspace \thinspace }h_u=g-\frac 1e\cdot g(u)\otimes g(u)\text{ ,}
\label{4.11} \\
g &=&g_{ij}\cdot dx^i.dx^j\text{ ,\thinspace \thinspace \thinspace
\thinspace }g_{ij}=g_{ji}\text{ \thinspace \thinspace \thinspace \thinspace
\thinspace ,\thinspace \thinspace \thinspace \thinspace \thinspace
\thinspace }\overline{g}=g^{ij}\cdot \partial _i.\partial _j\text{
\thinspace \thinspace \thinspace \thinspace \thinspace \thinspace
,\thinspace \thinspace \thinspace \thinspace \thinspace \thinspace
\thinspace \thinspace \thinspace }g^{ij}=g^{ji}\text{ \thinspace \thinspace .%
}  \nonumber
\end{eqnarray}

For $l_a:=0$, we have $\xi _{(a)}=\xi _{(a)\perp }$.

\section{Conclusion}

In the present paper we have recall some basic notions and mathematical
tools used in the structure of continuous media mechanics in Euclidean $E_n$ 
$(n=3)$, and (pseudo) Riemannian spaces $V_n$ $(n=4)$. Some of the notions
are generalized for $(\overline{L}_n,g)$-spaces. The presented results are
needed for the building out of continuous media mechanics in spaces with
contravariant and covariant affine connections and metrics. It should be
stressed that if we wish to describe a flow of material points (elements) we
could take into account the velocities $u$ and $\xi _{(a)}$ of the points
along the corresponding curves together with their kinematic characteristics
(such as relative velocity and acceleration, shear, rotation, and expansion
velocities and accelerations) which consideration are coming next.

\end{document}